\documentclass[%
 pra,
 twocolumn,
 amsmath,amssymb,
 aps,
]{revtex4-1}

\usepackage{graphicx}
\usepackage{dcolumn}
\usepackage{bm}
\usepackage{braket}
\usepackage{color}
\usepackage{hyperref}
\usepackage{booktabs,siunitx}
\usepackage{tikz}

\begin{document}

\title{Thermally induced entanglement of atomic oscillators}

\author{Pradip Laha}
 \email{pradip.laha@upol.cz}
\affiliation{%
 Department of Optics, Palack\'y University, 17. listopadu 1192/12, 77146 Olomouc, Czech Republic\\}%
\author{Luk{\'a}{\v s} Slodi{\v c}ka}%
 \email{slodicka@optics.upol.cz}
\affiliation{%
 Department of Optics, Palack\'y University, 17. listopadu 1192/12, 77146 Olomouc, Czech Republic\\}%
\author{Darren W. Moore}%
 \email{darren.moore@upol.cz}
\affiliation{%
 Department of Optics, Palack\'y University, 17. listopadu 1192/12, 77146 Olomouc, Czech Republic\\}%
\author{Radim Filip}%
 \email{filip@optics.upol.cz}
\affiliation{Department of Optics, Palack\'y University, 17. listopadu 1192/12, 77146 Olomouc, Czech Republic\\}%

\date{\today}

\begin{abstract}
Laser cooled ions trapped in a linear Paul trap are long-standing ideal candidates for realizing quantum simulation, especially of many-body systems. The properties that contribute to this also provide the opportunity to demonstrate unexpected quantum phenomena in few-body systems. A pair of ions interacting in such traps exchange vibrational quanta through the Coulomb interaction. This linear interaction can be anharmonically modulated by an elementary coupling to the internal two-level structure of one of the ions. Driven by thermal energy in the passively coupled oscillators, which are themselves coupled to the internal ground states of the ions, the nonlinear interaction autonomously and unconditionally generates entanglement between the mechanical modes of the ions. We examine this counter-intuitive thermally induced entanglement for several experimentally feasible model systems, and propose parameter regimes where state of the art trapped ion systems can produce such phenomena. In addition, we demonstrate a multiqubit enhancement of such thermally induced entanglement.

\end{abstract}

\keywords{
Jaynes-Cummings interaction, Beamsplitter, Nonclassicality, Logarithmic entanglement
}

\maketitle


\section{Introduction}
Quantum superposition phenomena are typically obtained by precisely and coherently driving nonlinear systems well-isolated from the environment at low temperatures. Quantum technologies taking advantage of such phenomena will be more practical if the quantum components instead operate autonomously~\cite{Bohr_Brask_2015}. To do so, they require resources such as quantum superposition and entanglement generated without an external coherent drive, and emerging purely from thermal energy. While counter-intuitive, as thermal energy is typically destructive to such resources, thermally induced entanglement can be generated quite straightforwardly. A prototypical example is a pair of two-level systems, one in the ground state and the other in a thermal state, which, through unitary interactions can exhibit greater entanglement for greater temperatures. Theoretical extensions of such cases to the interaction between a two-level system and an oscillator appeared some time ago~\cite{bose}, as also the case of entanglement between a pair of qubits in pure states induced by coupling to a thermal field~\cite{mskim}. Further studies show the manner in which separable pure states of pairs of oscillators or qubits generate \textit{enhanced} entanglement after being exposed either to a thermal environment~\cite{benatti_entangling_2006} or when interfaced directly with a single-mode thermal field~\cite{zhou_entanglement_2002,bradler_entanglement_2007,ping_continuous_2008}. Similar results hold for multiqubit collections of atoms~\cite{patrick_entangling_2011} as well. 
However, deterministic thermally enhanced entanglement arising from heating passively coupled oscillators with a Hilbert space of infinite dimension, coupled to cold atoms, remains hitherto unexplored. It is a nontrivial complementary task, as an excited two-level system only absorbs less than a single energy quantum from coupled thermal oscillators. 

With passive interactions, thermal oscillators do not entangle spontaneously. Therefore, oscillator-oscillator entanglement driven by thermal energy can be extremely useful. For example, conditional thermally induced entanglement can lead to Bell-inequality violations~\cite{filip02}. After many years of experimental development such experiments have become available using highly nonlinear controlled-swap interactions~\cite{yygao,hgan}. A simpler approach can be used to observe the thermally induced mutual coherence of atoms~\cite{filip14}. However, a deterministic version of thermally induced entanglement or mutual coherence for oscillators is still hard to imagine. An early step towards this intriguing possibility was the thermally induced unconditional generation of nonclassical oscillator states from the elementary Jaynes-Cummings (JC) interaction of the oscillator with a ground-state two-level system~\cite{slodicka,marek}. The nonclassicality emerges from a complex anharmonic modulation of the oscillator energy distribution by a simple, coherently interacting, two-level system initially in the ground state. In this method, the smallest nonlinear unit interacts with an oscillator to produce such phenomena. Using the thermal energy already present in the oscillator contrasts with the external environment used in reservoir engineering experiments~\cite{kienzler_quantum_2015}. Autonomous and unconditional generation of entanglement or coherence are the important resources to consider in scaling such behaviour to more complex systems. Here we focus on inducing entanglement in a pair of thermal oscillators using such methods.    

\begin{figure*}[ht!]
\centering
 \includegraphics[width=0.99\textwidth]{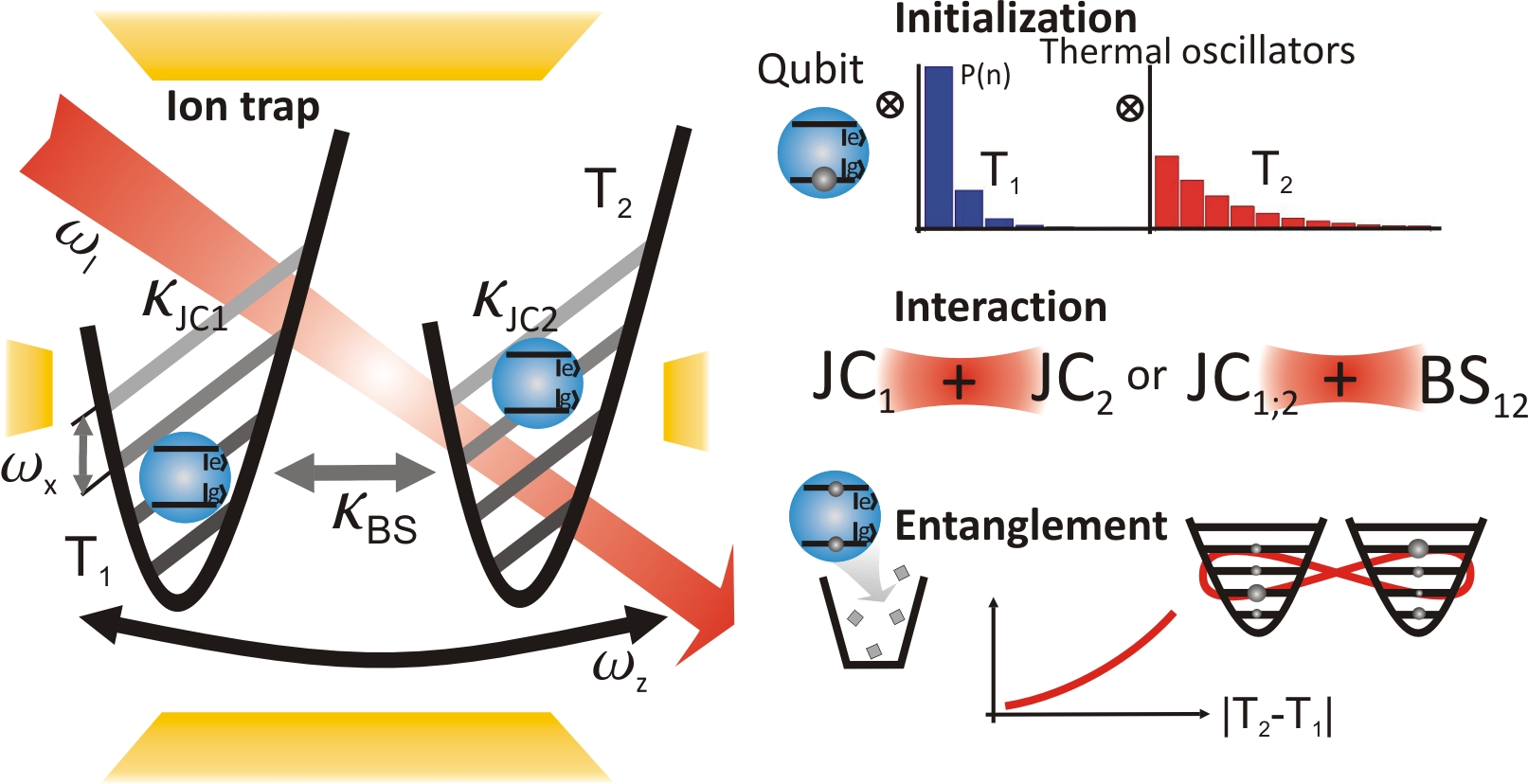}
\caption{A schematic of an ion trap setup that can realise the proposed deterministic generation of entanglement by increasing the temperature difference between two oscillator modes. Two ions in a linear trap align along the axial ($\omega_z$) direction and interact via the Coulomb force ($\kappa_\text{BS}$), sharing excitations among the radial vibrational modes. Alternatively, the internal energy of a single ion can be addressed in order to stimulate a Jaynes-Cummings interaction ($\kappa_\text{JC}$) between a pair of radial modes. A single qubit is prepared in the ground state, while the relevant radial modes (see main text) are prepared in initially uncorrelated thermal states. By selecting one of the interaction arrangements one of our models is implemented. From the full system dynamics of each model we discard the qubit contribution to calculate the entanglement of the joint two-field subsystem in terms of the logarithmic negativity. Note that system 1M (see text) is implemented with just a single ion with both radial modes coupled to the internal state of the ion, while system 1S requires the radial modes of two ions.}
 \label{fig0}
\end{figure*}

An array of laser driven ultracold ions confined in a linear Paul trap (see Fig.~\ref{fig0}) is an excellent and experimentally viable platform available for the implementation of a wide range of quantum protocols involving quantum simulation~\cite{johanning,blatt}, for example of Ising spin systems~\cite{friedenauer,kim1,kim2,islam}, the Bose-Hubbard (BH) model~\cite{zant,greiner,fisher} and the Jaynes-Cummings-Hubbard (JCH) model~\cite{greentree,hartmann,angelakis,rossini,irish,makin,toyoda1}. Recently, a single ion has been used to investigate quantum phase transitions in mechanical motion~\cite{mlcai_2021}. These systems possess several striking advantages such as the presence of long coherence times, the ability to efficiently manipulate the internal and motional (centre of mass) degrees of freedom, and individual access to each ion allowing implementation of quantum gates between specific ions~\cite{cirac,toyoda2}. Spontaneous emission of the internal states and damping of the motion of the ions are effectively minimised in state of the art experiments by tuning the appropriate metastable transitions of the ions stored in ultrahigh precision spectroscopy and maintaining low pressures for long times such that collisions with background atoms and other incoherent couplings are negligible~\cite{leibfried03}. The high level of control afforded by such systems combined with their anticipated scalability in simulating many-body effects provide us with a framework in which the autonomous quantum behaviour that may become a necessity of larger-scale quantum technologies can be investigated. While we focus entirely on trapped ion systems in our analysis, we also note the potential application of our scheme to superconducting circuits~\cite{hays_continuous_2020} which share many properties we make use of below.

\section{\label{sec:results} Thermally induced entanglement}
A collection of ions can be confined in three dimensions by a trap that is harmonic at the level of the motion of the ions. That is, the ions are confined by harmonic potentials of frequencies $\omega_\alpha$, $\alpha=x,y,z$. If $\omega_z\ll\omega_{x,y}$ then the ions form a chain by aligning themselves along the axial ($z$) direction, with equilibrium positions $(0,0,z^0_i)$. The vibrational motion of the ions occurs around equilibrium points situated along this chain. 

\begin{figure*}[ht!]
\centering
\begin{tikzpicture}

 \draw [ultra thick,dashed,draw=black, fill=orange, fill opacity=0.2] (-2,2) ellipse (0.65cm and 0.7cm);
 \draw [very thick, draw=black] (-2.50,2.35) -- (-1.50,2.35);
 \draw [very thick, draw=black] (-2.48,1.65) -- (-1.50,1.65);
 \node at (-2.0, 2.52){\scriptsize $\ket{e}$};
 \node at (-2.0, 1.48){\scriptsize $\ket{g}$};
 \draw [<->, dashed, thick, draw=black] (-2.0,1.65) -- (-2.0,2.35) node[pos=0.5,left] {\footnotesize$\omega$} ;
 
 \draw [<->, very thick, draw=black] (-1.32,2.0) -- (-0.34,2.0) node[pos=0.5,below] {\footnotesize$\kappa_{\textrm{JC}_1}$} node[pos=0.5,above] {\footnotesize JC};
 
 \draw [ultra thick,dashed,draw=black, fill=purple, fill opacity=0.2] (0.35,2) ellipse (0.65cm and 0.7cm);
 \draw [very thick, draw=black] (-0.125,2.40) parabola[parabola height=-1.02cm] (0.82,2.40);
 \draw [thick, draw=black] (0.18,1.50) -- (0.52,1.5);
 \draw [thick, draw=black] (0.08,1.75) -- (0.62,1.75);
 \draw [thick, draw=black] (-0.03,2.00) -- (0.73,2.00);
 \draw [thick, draw=black] (-0.10,2.25) -- (0.80,2.25);
 \node at (0.35,2.45){\scriptsize$a_{1},\, a_{1}^{\dagger}$};

 \draw [<->, very thick, draw=black] (1.04,2.0) -- (2.02,2.0) node[pos=0.5,below] {\footnotesize$\kappa_{\textrm{BS}_{12}}$} node[pos=0.5,above] {BS};

 \draw [ultra thick,dashed,draw=black, fill=yellow, fill opacity=0.2] (2.7,2) ellipse (0.65cm and 0.7cm);
  \draw [very thick, draw=black] (2.22,2.4) parabola[parabola height=-1.02cm] (3.18,2.4);
 \draw [thick, draw=black] (2.52,1.50) -- (2.85,1.50);
 \draw [thick, draw=black] (2.40,1.75) -- (2.97,1.75);
 \draw [thick, draw=black] (2.32,2.00) -- (3.07,2.00);
 \draw [thick, draw=black] (2.25,2.25) -- (3.16,2.25);
 \node at (2.71,2.45){\scriptsize$a_{2},\, a_{2}^{\dagger}$};

\end{tikzpicture}\\ 
\vspace{2ex}
\includegraphics{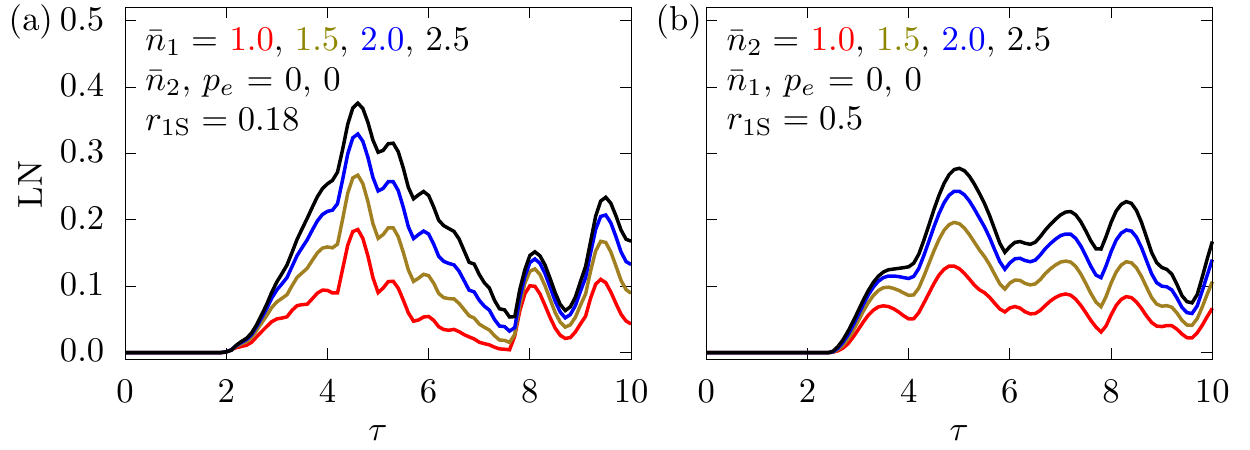}\\
\vspace{2ex}
\begin{tikzpicture}

 \draw [ultra thick,dashed,draw=black, fill=purple, fill opacity=0.2] (5.8,2) ellipse (0.65cm and 0.7cm);
 \draw [very thick, draw=black] (5.32,2.4) parabola[parabola height=-1.0cm] (6.27,2.4);
 \draw [thick, draw=black] (5.63,1.50) -- (5.96,1.50);
 \draw [thick, draw=black] (5.50,1.75) -- (6.09,1.75);
 \draw [thick, draw=black] (5.41,2.00) -- (6.18,2.00);
 \draw [thick, draw=black] (5.35,2.25) -- (6.25,2.25);
 \node at (5.80,2.45){\scriptsize$a_{1},\, a_{1}^{\dagger}$};

\draw [<->, very thick, draw=black] (6.47,2.0) -- (7.45,2.0) node[pos=0.5,below] {\footnotesize$\kappa_{\textrm{JC}_1}$} node[pos=0.5,above] {JC};

\draw [<->, very thick, draw=black] (8.82,2.0) -- (9.8,2.0) node[pos=0.5,below] {\footnotesize$\kappa_{\textrm{JC}_2}$} node[pos=0.5,above] {JC};

 \draw [ultra thick,dashed,draw=black, fill=orange, fill opacity=0.2] (8.14,2) ellipse (0.65cm and 0.7cm);
 \draw [very thick, draw=black] (7.63,2.35) -- (8.66,2.35);
 \draw [very thick, draw=black] (7.63,1.65) -- (8.66,1.65);
 \node at (8.15, 2.52){\scriptsize $\ket{e}$};
 \node at (8.15, 1.48){\scriptsize $\ket{g}$};
 
 \draw [<->, dashed, thick, draw=black] (8.15,1.65) -- (8.15,2.35) node[pos=0.5,left] {\footnotesize$\omega$} ;

 \draw [ultra thick,dashed,draw=black, fill=yellow, fill opacity=0.2] (10.49,2) ellipse (0.65cm and 0.7cm);
  \draw [very thick, draw=black] (10.01,2.4) parabola[parabola height=-1.0cm] (10.97,2.4);
 \draw [thick, draw=black] (10.32,1.50) -- (10.66,1.50);
 \draw [thick, draw=black] (10.22,1.75) -- (10.76,1.75);
 \draw [thick, draw=black] (10.10,2.00) -- (10.85,2.00);
 \draw [thick, draw=black] (10.05,2.25) -- (10.92,2.25);
  \node at (10.5,2.45){\scriptsize$a_{2},\, a_{2}^{\dagger}$};

\end{tikzpicture}\\ 
\vspace{2ex}
\includegraphics{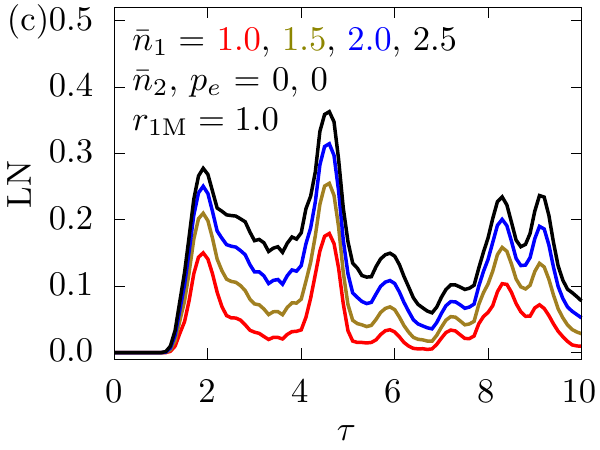}\\
\caption{Temporal evolution of thermally induced entanglement by a single qubit. The systems embodied by Hamiltonians $H_{\mathrm{1S}}$ and $H_{\mathrm{1M}}$ are shown schematically above their respective entanglement dynamics. (a)-(b) The logarithmic negativity (LN) as a function of the scaled time $\tau$ for various values of $\bar{n}_1$ ($\bar{n}_2$), while $\bar{n}_2=0$ ($\bar{n}_1=0$). The qubit is initially in the ground state, while the coupling ratios are given by $r_\mathrm{1S} = 0.18$ (0.5). After a short latent period entanglement becomes visible, and increasing the temperature difference significantly increases the LN. (c) The logarithmic negativity for $H_{\mathrm{1M}}$ with coupling ratio  $r_\mathrm{1M} = 1$. The symmetry of $H_{\mathrm{1M}}$ ensures that no different behaviour is observed by swapping the roles of $\bar{n}_1$ and $\bar{n}_2$ whereas the asymmetry of $H_{\mathrm{1S}}$ is visible by the reduced amount of entanglement produced when the thermal energy is concentrated in the oscillator distant from the qubit. This asymmetry also modifies the optimal coupling strength, requiring stronger coupling in order to maximise the induced entanglement when the thermal noise is initialised in oscillator 2.}
 \label{fig1}
\end{figure*}
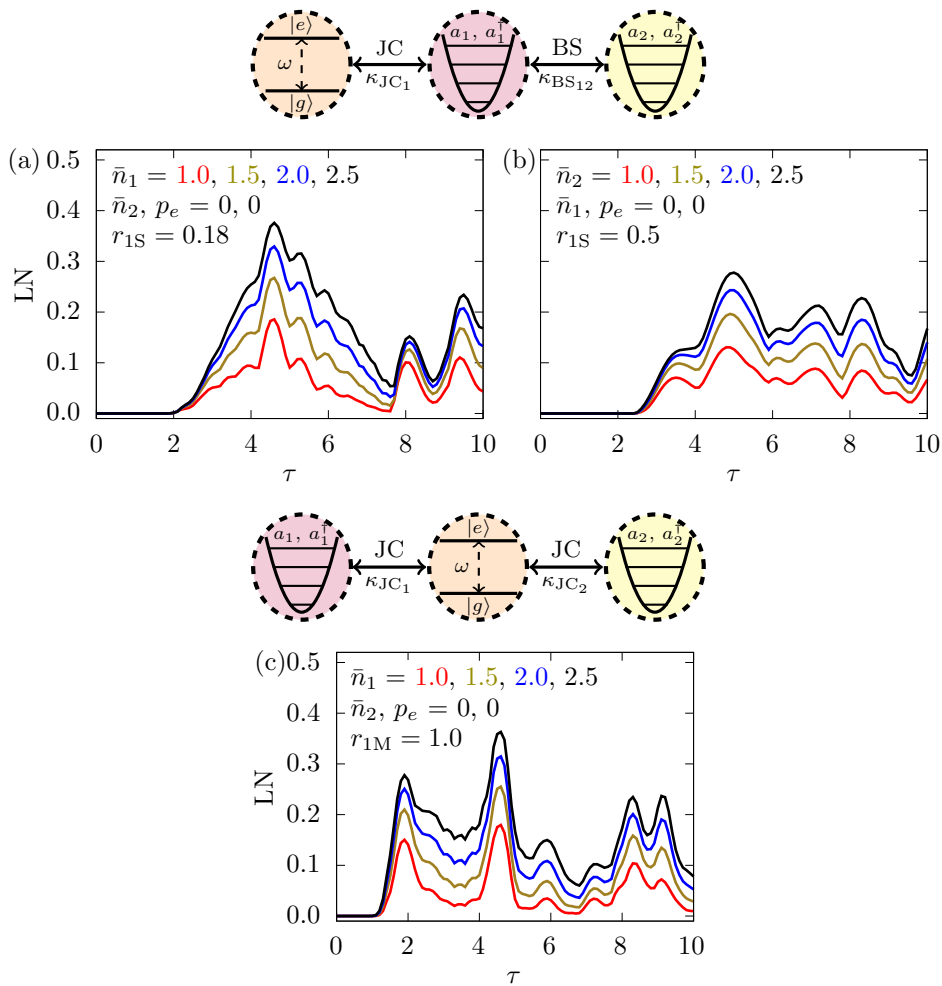

In the following derivation we focus on a qubit interacting with a pair of oscillator modes. In the case of a beamsplitter interaction two ions are required, however we emphasise that a single ion can be leveraged to implement an alternate configuration involving only Jaynes-Cummings interactions (see schematics of Fig.~\ref{fig1}). A pair of confined ions undergo vibrational motion in the radial directions $x$ and $y$. The Coulomb interaction can induce the hopping of vibrational quanta, phonons, between the corresponding radial modes. This can straightforwardly be seen through a Hamiltonian analysis of the system.
The generic Hamiltonian including both the trapping potential and the Coulomb interaction between the ions is given by
\begin{equation}
  H = \frac{1}{2m}\sum_{\alpha=x,y,z} \sum_{i=1}^{2} p_{\alpha,i}^{2} + \sum_{\alpha=x,y,z} \frac{1}{2} m \omega_{\alpha}^2 \sum_{i=1}^{2} \alpha_{i}^{2} + \frac{e^{2}}{\left\vert \mathbf{r}_{1} - \mathbf{r}_{2}\right\vert}\,,
  \label{eqn:eqn1}
\end{equation}
where $\alpha_i$ and $p_{\alpha,i}$ are respectively the $\alpha$ components of the displacement from the equilibrium position and the momentum of the $i$th ion, and $m$ and $e$ are respectively the mass and charge of an ion, while the $\mathbf{r}_i=(x_i,~  y_i,~  z_i)^\top$ are the positions of the ions relative to the equilibrium position. The distance term in the Coulomb potential can be expanded around the displacements from equilibrium to include up to harmonic terms. This approximation leads to the decoupling of the individual modes which is a good approximation for axial potential frequencies much smaller than radial ones and for small thermal energies of ions. Now, decomposing the dynamics into one of the transverse directions, say $x$, we have the Hamiltonian~\cite{porras,deng} (see the Supplementary Material for a detailed derivation)
\begin{equation} 
  H = \frac{1}{2m} \sum_i p_{i}^{2} + \frac{1}{2} m  \sum_{i}\omega_{x_i}^2 x_{i}^{2} + \frac{e^2}{2} \frac{\left(x_1-x_2\right)^{2}}{\left\vert z^{0}_{1} - z^{0}_{2}\right\vert^{3}}\,.
  \label{eqn:eqn2}
\end{equation}
The Coulomb effect therefore induces frequency shifts in the oscillators, $\omega_x^2\rightarrow\omega_x^2+\frac{e^2}{2m\vert z^0_1-z^0_2 \vert^3}$. With this the Hamiltonian  can be written in second quantization by promoting $x_i$ and $p_i$ to dimensionless quadrature operators $\hat{x}_i = \sqrt{\frac{\hbar}{2m\omega_x}}(a_i+a_i^\dagger)$ and $\hat{p}_i = \sqrt{\frac{\hbar m\omega_x}{2}}i(a_i^\dagger - a_i)$. Neglecting the zero-point energy of the oscillators, we may write
\begin{equation}
  H=\hbar\,\omega_x\sum_ia^\dagger_{i}a_i+\kappa\left(a_1+a_1^\dagger\right)\left(a_2+a_2^\dagger\right)\,,
  \label{eqn:eqn3}
\end{equation}
where $\kappa = -\frac{\hbar e^2}{2m\omega_x \vert z^0_1-z^0_2\vert^3}$. In a frame rotating with respect to the free motion, the interaction Hamiltonian is composed of resonant and fast-rotating terms. In the rotating wave approximation we retain only the resonant terms which constitute a beamsplitter-like interaction allowing excitation transfer between the modes,
\begin{equation}
    H_{\text{BS}_{ij}}=\kappa_{\text{BS}_{ij}}\left(a_i^\dagger a_j+a_ia_j^\dagger\right)~ \quad (i\ne j)\,.
  \label{eqn:eqn4}
\end{equation}

Simultaneously, if one considers the internal energy levels of an ion, addressed by an external electromagnetic wave, then a discrete-variable nonlinearity can be introduced~\cite{wineland}. Addressing a pair of energy levels leads to an interaction Hamiltonian in the rotating wave approximation of the form
\begin{equation}
    H_\text{int}=\hbar\,\Omega\,\sigma_+e^{i\left(\eta\left(a_i+a_i^\dagger\right)-\delta t+\phi\right)}+\text{h.c.}\,,
  \label{eqn:eqn7}
\end{equation}
where $\delta$ is the detuning of the driving from the frequency of the two-level system and $\Omega$ is the coupling strength.
Assuming the Lamb-Dicke approximation corresponding to $\eta\sqrt{\braket{x_i}^2}<<1$ and setting the laser to the resonance with the first red motional sideband, $\delta=-\omega_x$, gives
\begin{equation}
    H_\text{int}=\kappa_{\text{JC}_i}
    \left(\sigma_{+} a_{i} e^{i\phi} + \sigma_{-} a_{i}^{\dagger} e^{-i\phi}\right)\,.
  \label{eqn:eqn8}
\end{equation}
where $\kappa_{\text{JC}_i}=\hbar\,\Omega\,\eta$. A coherent JC coupling to the second radial mode can simultaneously be implemented with a bichromatic laser excitation that addresses two red sidebands simultaneously.

While the oscillator interaction is passively linear, one may take advantage of the nonlinear coupling inherent in the JC model in order to use the temperature difference of the oscillator pair to induce entanglement. Various configurations can be imagined and we analyse and contrast two basic instances below. 

 The entanglement production is driven by the difference between the initial temperatures of the subsystems. Thermal states of the oscillator and qubit, in thermal equilibrium with nonzero temperatures $T^{\text{O}}_i$ and $T^{\text{q}}$ respectively, are characterised by the Boltzmann distributions
\begin{align}
 \rho^{\text{O}}_{i} &= \sum_{n_{i}=0}^{\infty} \frac{\bar{n}_{i}^{n_{i}}}{\left(1+\bar{n}_{i}\right)^{n_{i}+1}}  \ket{n_{i}}\bra{n_{i}},\\
 \rho^{\text{q}}_{j}  &= p_{e_j} \ket{e}\bra{e} + \left(1 - p_{e_j}\right) \ket{g}\bra{g}\,,
\end{align}
where $\ket{g}$ and $\ket{e}$ are the energy eigenstates of the qubit. The mean excitation $\bar{n}_{i}$ of the $i$th oscillator is related to $T^{O}_{i}$ via 
$\bar{n}_{i} = \left(e^{\frac{\hbar\omega_x}{k_{B} T^{\text{O}}_{i}}} - 1\right)^{-1}$,
where $k_{\text{B}}$ is the Boltzmann constant.
The probability of finding the qubit in the excited state is
$p_{e_j} = \frac{e^{-\frac{\hbar\omega_\text{int}}{k_{B} T^{\text{q}}_{j}}}}{ e^{-\frac{\hbar\omega_\text{int}}{k_{B} T^{q}_{j}}}  + 1}$.
In trapped ion systems, the characteristic decoherence rates are small compared to the short interaction times required to generate entanglement~\cite{leibfried03}, and hence we assume unitary dynamics (see, however, the Supplementary Material). 

\subsection{\label{subsec:singlequbit} Thermally induced entanglement via a single qubit}

The simplest possible generalisation of the thermally induced entanglement protocols involves a pair of oscillator modes and a single qubit. The two elementary  arrangements of these three systems involve the qubit coupled to one member of the oscillator mode pair which is coupled to the other member via the beamsplitter interaction, or a single qubit acting as mediator between a pair of uncoupled oscillator modes (Fig.~\ref{fig1} schematics). The interaction Hamiltonians of such systems can be expressed as
\begin{align}
H_{\mathrm{1S}} &= H_{\text{JC}_1} + H_{\text{BS}_{12}}, \label{eqn:ham_1}\\
H_{\mathrm{1M}} &= H_{\text{JC}_1} + H_{\text{JC}_2}\,. \label{eqn:ham_2}
\end{align}
Here $\mathrm{S}$ ($\mathrm{M}$) stands for a qubit at the side (in the middle). $H_{\mathrm{1S}}$, in taking advantage of the beamsplitter interaction requires the interaction of the radial modes of two ions, whereas $H_{\mathrm{1M}}$ requires a single qubit and is implemented using the radial modes of a single ion. To study the entanglement dynamics we measure the entanglement between the reduced state $\rho$ of the oscillator pair by numerically estimating the logarithmic negativity (LN)~\cite{vidal,plenio}, which is defined by
\begin{equation}
    \text{LN}=\log_2||\rho^{\top_A}||_1\,,
\end{equation}
where $\top_A$ is the partial transpose operation and $||\rho||_1$ denotes the trace norm of the operator $\rho$.

Contrary to intuition, increasing the temperature difference of the oscillator modes enhances the entanglement produced by the unitary evolution of these systems.  
We consider these systems in turn below. In discussing system $\mathrm{1S}$ it is useful to first define a single timescale between those of $\kappa_{\text{JC}_1}$ and $\kappa_{\text{BS}_{12}}$ by introducing $r_\mathrm{1S}=\frac{\kappa_{\text{BS}_{12}}}{\kappa_{\text{JC}_1}}$ such that the dynamics occurs in units of $\tau = \kappa_{\text{JC}_1}t$. For system $\mathrm{1M}$ we write $r_\mathrm{1M}=\frac{\kappa_{\text{JC}_2}}{\kappa_{\text{JC}_1}}$. The symmetry in $H_{\mathrm{1M}}$ means that $r_\mathrm{1M}^{-1}$ simply represents a relabelling of the oscillators. It is possible, without loss of generality, to study only the effects of $p_e$ and $\bar{n}_2$ with respect to $\bar{n}_1$ with $0<r_{\mathrm{1M}}\le1$. 


We first consider $p_{e} = 0$, so that the atom is in its ground state and  examine the cases where either $\bar{n}_{1} = 0$ or $\bar{n}_{2} = 0$ with the remaining oscillator having a non-zero mean phonon number. Figs.~\ref{fig1}~(a-c) illustrate the generation of LN over scaled time $\tau$ for different initial distributions of the thermal noise in the first and second oscillators. Firstly, for system $\mathrm{1S}$ (top), we note that the general behaviour indicates that thermal noise in the first oscillator, the one directly coupled to the qubit, is most significant for generating entanglement. This is supported by the noise distribution in favour of oscillator 1 producing much more entanglement with a much weaker qubit-oscillator coupling rate $r_{\mathrm{1S}}$, compared to the initial thermal noise in oscillator 2. For system $\mathrm{1M}$ (bottom) the symmetry of the configuration means that the two variations are identical. The LN is noticeably less than that of system $\mathrm{1S}$. For both systems the entanglement detected by the LN emerges after a short delay, and then experiences periodic collapses and revivals. This delay indicates that short time approximations of the JC interaction will not produce nonzero LN of the oscillators and that this is instead a property of the full complex dynamics introduced by the qubit. Interestingly, the first rise in entanglement for system $\mathrm{1M}$ takes approximately half the time of system $\mathrm{1S}$. Increasing the thermal energy increases the entanglement generated, but notably does not change the profiles of the collapses and revivals, which are monotonic with respect to the thermal noise, up to saturation; when the temperature becomes too high, the entanglement saturates. 

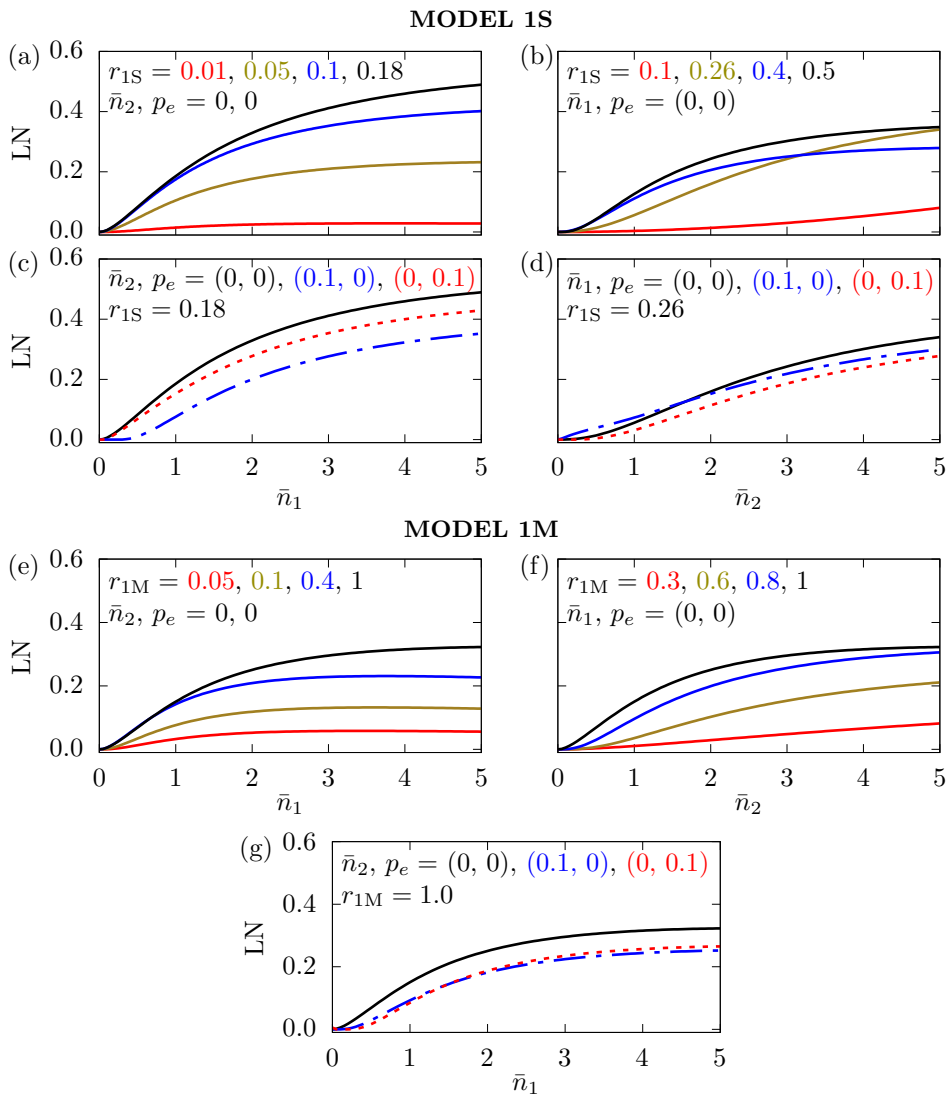
\begin{figure*}[ht!]
\begin{center}
\begin{tikzpicture}
\node at (9.3,0.0){\bf MODEL 1S};
\end{tikzpicture}\\ \vspace{-0.5ex}
\includegraphics{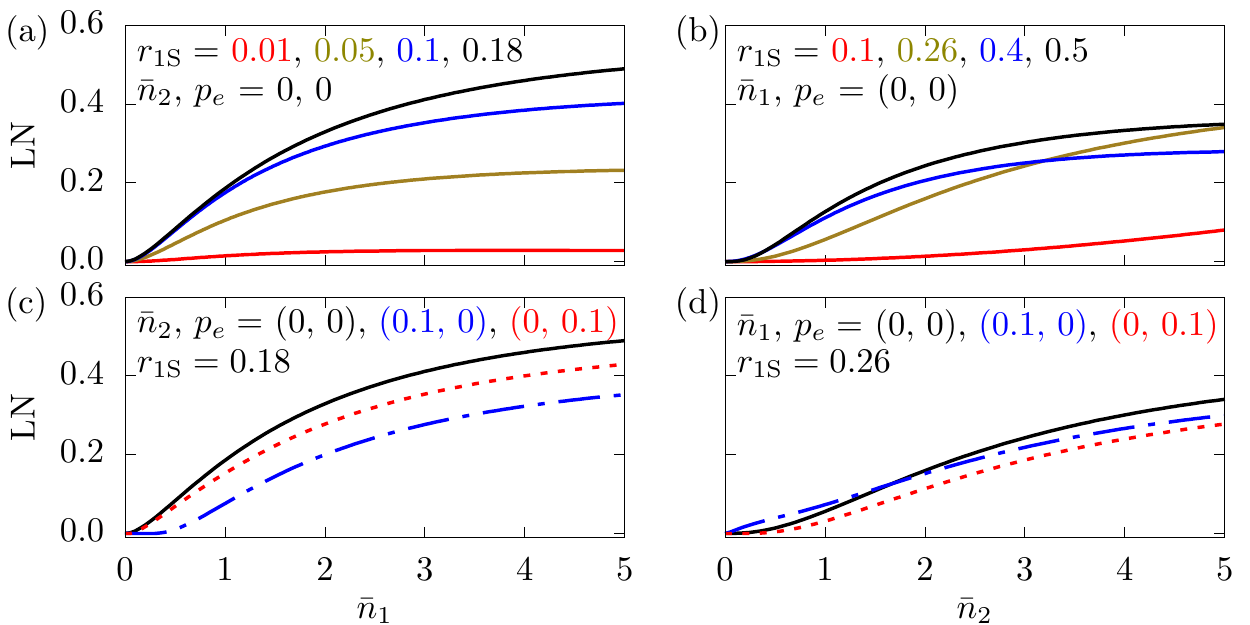} \\ \vspace{-0ex}
\begin{tikzpicture}
\node at (6.3,0.0){\bf MODEL 1M};
\end{tikzpicture}\\ \vspace{0ex}
\includegraphics{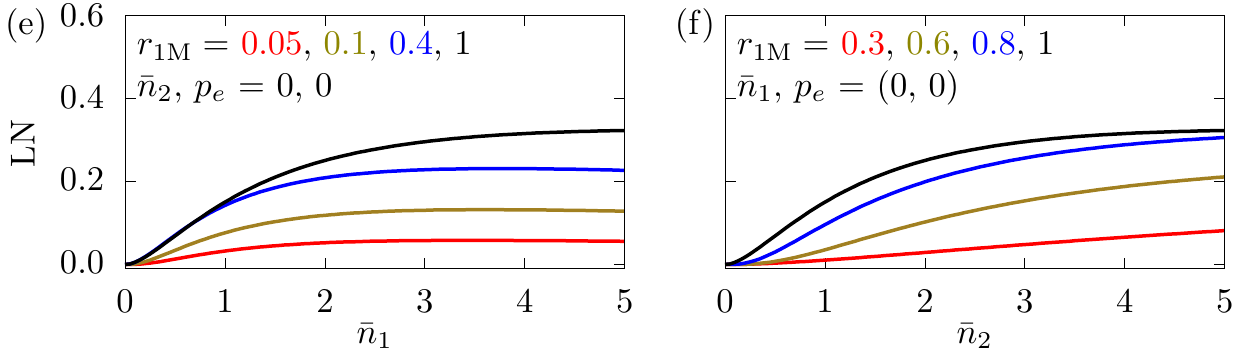} \\ \vspace{1ex}
\includegraphics{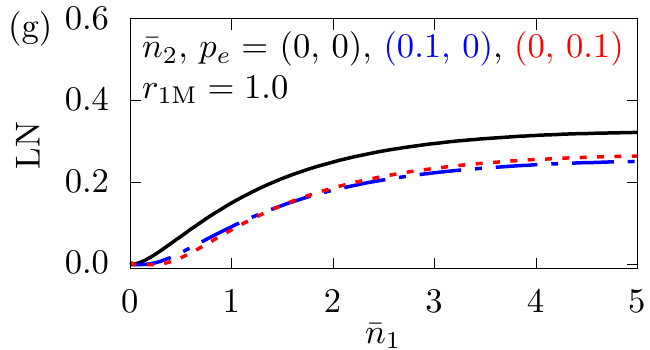} \\ \vspace{-1ex}
\caption{Entanglement production from $H_{\mathrm{1S}}$ (a-d) and $H_{\mathrm{1M}}$ (e-g) in terms of LN at optimal time  as a function of $\bar{n}_1$ and $\bar{n}_2$. Asymmetrically increasing the thermal noise of one oscillator increases the entanglement up to a saturation point. Additionally, increasing the coupling $r_\mathrm{1S}$ or $r_\mathrm{1M}$ up to a critical value increases the entanglement (black lines). Notably, $H_{\mathrm{1M}}$ has a critical value at the symmetric value $r_\mathrm{1M}=1$.  The effects of increasing the mixedness of the remaining subsystems for the critical values of $r_\mathrm{1S}$ (c-d) and $r_\mathrm{1M}$ (g) result in a decrease in entanglement generation. $H_{\mathrm{1S}}$ generates relatively less entanglement (b) and requires larger coupling to achieve it. Additionally, it shows crossings in LN that break the pattern of monotonic increase (with regard to $r_\mathrm{1S}$) seen otherwise. Panel (d) shows the importance of the thermal noise directly interfacing with the nonlinearity since, in contrast to the typical entanglement dynamics, adding thermal noise to oscillator 1 (decreasing the temperature difference) increases the entanglement for small $\bar{n}_2$.}
\label{fig2}
\end{center}
\end{figure*}

In what follows we select the optimal evolution time as the first maximum in the entanglement dynamics to analyse the intermediate parameter regimes. In some cases (Fig.~\ref{fig1}~(b), for example), a later maximum can be higher than the first but we wish to minimise any potential detrimental effects from decoherence of the system. Figs.~\ref{fig2}~(a-b) and Figs.~\ref{fig2}~(e-f) illustrate the manner in which entanglement increases with an increase in the thermal noise. Regardless of the coupling strengths, increasing the thermal noise increases the entanglement, up to saturation and up to a critical value (black line) above which entanglement generation begins to be inhibited again. The second and fourth rows indicate that increasing the mixedness in the remaining components of both systems decreases the capacity to generate entanglement. In fact, if $\bar{n}_1=\bar{n}_2$ there is no entanglement generation even with the qubit in the ground state. Both the ground state of the qubit and the temperature difference are important for entanglement generation. As said however, thermal noise in oscillator 2 for system $\mathrm{1S}$ is less effective in generating entanglement, as seen in  panels (b) and (d), so that while the temperature difference drives the entanglement generation, the distribution of the thermal noise must also be such that it interacts directly with the qubit induced nonlinearity. This is supported by panel (d) which shows that adding thermal noise to the more effective oscillator 1 does not increase the entanglement, but rather is consistent with our interpretation that it is the temperature difference that drives the entanglement generation with one exception. Fig.~\ref{fig2}~(d) shows that for very small values of $\bar{n}_2$, increasing $\bar{n}_1$ is useful. This supports the argument that it is both the temperature difference and the direct interfacing of the nonlinearity with the thermal energy that are important for entanglement generation. Larger values of $\bar{n}_2$ obviate this effect. In contrast to the other results, panel (b) shows a crossing in the LN. Similar to increasing $\bar{n}_1$, strong coupling can compensate for the incorrect distribution of the thermal noise for lower values of $\bar{n}_2$. The symmetry of $\mathrm{1M}$ means that when the roles of oscillators 1 and 2 are swapped the results do not change. This is further supported by the fact that the optimal coupling ratio is the balanced $r_\mathrm{1M}=1$. 

\subsection{Experimental feasibility}

Experimental progress in the last decade towards implementing the BH or JCH model for trapped ions has resulted in the observation of phonon-hopping between radial modes of two trapped ions~\cite{haze2012,toyoda2015} or between the pair of radial modes in a single ion~\cite{zhang2018}. The significance of such models is typically held to lie in their many-body qubit properties such as M\"{o}tt insulator and superconducting phases~\cite{porras,toyoda1}. Such nonlinear interactions along with the ability to directly address specific ions also allow a collection of trapped ions to act as a quantum simulator. In this article, we have rather focused on applying these developments to a small-scale few-body quantum simulator of nonlinearly induced effects in linear oscillators. A similar effort can be seen in a recent experiment on quantum phase transitions with a single ion~\cite{mlcai_2021}. These experiments are well-suited to investigate the autonomously generated entanglement we have studied here. 

In particular, system $\mathrm{1S}$ has already seen experimental realization in~\cite{haze2012} and~\cite{toyoda2015}, demonstrating the hopping of excitations and the Hong-Ou-Mandel effect respectively. Two ${}^{40}\text{Ca}^+$ ions, confined axially by a DC electric field and radially by radio frequency electric fields, share radial mode phonons mediated by the Coulomb interaction. The axial frequency is engineered to be much lower than the radial frequencies, so that the ions align. An extra DC field is applied to lift the degeneracy of the two radial modes, which can then be addressed distinctly. The internal state of one of the ions can be addressed by external lasers, initiating the JC interaction and a pulse on the blue Rabi sideband can prepare the qubit in the ground state.

In system $\mathrm{1S}$, the largest entanglement generation occurs for $r_\mathrm{1S}=0.18$. In Ref.~\cite{haze2012}, the ratio of such a coupling was $r_\mathrm{1S}\simeq0.32$, which is well within the range of our results. However, since this is above the optimal value for $\bar{n}_1$, it might be more effective to place the thermal energy in the second oscillator, with a nonzero $\bar{n}_2$. This corresponds more closely to the gold line in panel (b) of Fig.~\ref{fig2}. Alternatively, the ratio can be further decreased to the optimal values by increasing the inter-ion distance $|z_1^0-z_2^0|$. This is mainly controlled by the axial frequency $\omega_z$. If this can be reduced while maintaining the conditions required for the approximations of the model then the more effective regimes of entanglement generation may be achievable. To realize the initial temperature difference in the oscillator pair, Doppler and sideband cooling can be applied to cool the vibrational modes towards the ground states, reaching average phonon numbers as low as $\bar{n}=0.08$. For thermally induced entanglement, it is the temperature difference that matters, and hence cooling of one of the two vibrational modes is sufficient to institute the required initial state. It is possible to cool only one of the modes since the ions are individually addressable. However, the cooling time is typically a substantial proportion of the evolution time. For this reason, we recommend that the initial states of the ions are prepared with a large inter-ion distance, then brought together, so that the excitation transfer dynamics only begins once the ions are prepared. In our results we have optimised the evolution time which usually requires only a few Rabi oscillations to produce nontrivial correlations. Therefore the entanglement is generated on a timescale in which dissipative effects are typically negligible. Notwithstanding this condition we give some attention to important decoherence effects in the Discussion and the Supplementary Material.

Alternatively, system $\mathrm{1M}$ is implemented in 
Ref.~\cite{zhang2018} in order to study NOON states. In this example, a single ${}^{171}$Yb$^+$ ion is confined in a similar trap and an internal hyperfine transition is addressed to couple it to the radial motion. In this experiment the ratio takes the value $r_\mathrm{1M}\simeq1$, replicating our optimal coupling conditions. Similar remarks regarding the coherence times and state preparation apply here.

\subsection{Multi-qubit thermal enhancement of entanglement}

The JC model is readily generalised to the Tavis-Cummings model~\cite{tavis_exact_1968,tavis_approximate_1969}, in which a set of identical two-level systems interact with a single oscillator mode, and without a rotating wave approximation generalises to the Dicke model~\cite{kirton_introduction_2019,aedo_analog_2018}. Such systems exhibit phase transitions that amount to collective enhancement of coherent quantum phenomena such as lasing and superradiance~\cite{kirton_superradiant_2018}. Motivated by such examples we briefly extend our analysis to demonstrate the potential enhancement of thermally induced entanglement via the coherent action of multiple qubits. Since we are interested only in demonstrating the existence of such an enhancement and the number of possible configurations of the system increases combinatorially with the number of qubits, we restrict ourselves to an analysis of two qubits interacting with an oscillator pair. Physically such systems can be conceived of as addressing a second ion with a driving electromagnetic field, inducing the JC interaction between this ion and the vibrational mode involved in the Coulomb-induced excitation exchange. Such a setup can be envisioned as involving a second narrow-waist laser directed at the second ion, or a broad-waist laser that addresses both ions, simultaneously exciting the internal motion. We again assume a ground state of the qubit, and analyse several possible configurations of the ion-oscillator interactions. 

\begin{figure*}[ht!]
\centering
\begin{tikzpicture}

\node(label) at (0.6,2.3){\bf MODEL 2S};

 \draw [ultra thick,dashed,draw=black, fill=orange, fill opacity=0.2] (-1.5,1.25) ellipse (0.65cm and 0.7cm);
 \draw [very thick, draw=black] (-2.0,0.92) -- (-1.0,0.92);
 \draw [very thick, draw=black] (-2.0,1.60) -- (-1.0,1.60);
 \node at (-1.5, 1.76){\scriptsize $\ket{e}$};
 \node at (-1.5, 0.75){\scriptsize $\ket{g}$};

 \draw [<->, very thick, draw=black] (-0.82,1.40) -- (0.22,0.55);
 \node at (-0.46,0.8) {\footnotesize JC};
 \node at (+0.00,1.1)  {\scriptsize$\kappa_{\text{JC}_{a1}}$};

 \draw [<->, dashed, thick, draw=black] (-1.5,0.92) -- (-1.5,1.6) node[pos=0.5,left] {$\omega$} ;
 
 \draw [ultra thick,dashed,draw=black, fill=orange, fill opacity=0.2] (-1.5,-0.75) ellipse (0.65cm and 0.7cm);
 \draw [very thick, draw=black] (-2.0,-0.45) -- (-1.0,-0.45);
 \draw [very thick, draw=black] (-2.0,-1.10) -- (-1.0,-1.1);
 \node at (-1.5, -0.25){\scriptsize $\ket{e}$};
 \node at (-1.5, -1.25){\scriptsize $\ket{g}$};

 \draw [<->, very thick, draw=black] (-0.80,-0.75) -- (0.20,-0.00) ;
 \node at (-0.45,-0.20) {\footnotesize JC};
 \node at (-0.0,-0.6)  {\footnotesize $\kappa_{\text{JC}_{b1}}$};

 \draw [<->, dashed, thick, draw=black] (-1.5,-1.1) -- (-1.5,-0.45) node[pos=0.5,left] {$\omega$} ;

 \draw [ultra thick,dashed,draw=black, fill=purple, fill opacity=0.2] (0.85,0.25) ellipse (0.65cm and 0.7cm);
 \draw [very thick, draw=black] (0.38,0.65) parabola[parabola height=-1.0cm] (1.3,0.65);
 \draw [thick, draw=black] (0.7,-0.25) -- (0.97,-0.25);
 \draw [thick, draw=black] (0.58,0.00) -- (1.12,0.00);
 \draw [thick, draw=black] (0.5,0.25) -- (1.18,0.25);
 \draw [thick, draw=black] (0.4,0.50) -- (1.25,0.50);
 \node at (0.82,0.7){\scriptsize$a_{1},\, a_{1}^{\dagger}$};

 \draw [<->, very thick, draw=black] (1.53,0.25) -- (2.81,0.25) node[pos=0.5,below] {\footnotesize$\kappa_{\text{BS}_{12}}$} node[pos=0.5,above] {BS};

 \draw [ultra thick,dashed,draw=black, fill=yellow, fill opacity=0.2] (3.5,0.25) ellipse (0.65cm and 0.7cm);
 \draw [very thick, draw=black] (3.02,0.65) parabola[parabola height=-1.0cm] (3.97,0.65);
 \draw [thick, draw=black] (3.35,-0.25) -- (3.64,-0.25);
 \draw [thick, draw=black] (3.22,0.00) -- (3.78,0.00);
 \draw [thick, draw=black] (3.12,0.25) -- (3.85,0.25);
 \draw [thick, draw=black] (3.04,0.50) -- (3.95,0.50);
 \node at (3.46,0.7){\scriptsize$a_{2},\, a_{2}^{\dagger}$};
\end{tikzpicture} \\
\includegraphics{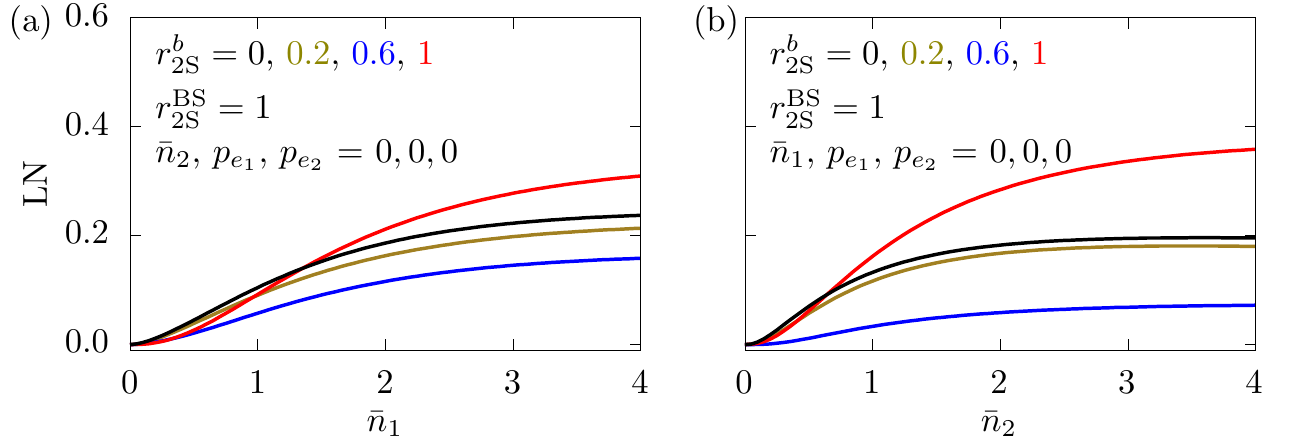} \\ \vspace{2ex}
\begin{tikzpicture}
 \node(label) at (9.1,2.3){\bf MODEL 2M};
\draw [ultra thick,dashed,draw=black, fill=purple, fill opacity=0.2] (6.4,0.25) ellipse (0.65cm and 0.7cm);
 \draw [very thick, draw=black] (6.87,0.65) parabola[parabola height=-1.0cm] (5.92,0.65);
 \draw [thick, draw=black] (6.56,-0.25) -- (6.24,-0.25);
 \draw [thick, draw=black] (6.68, 0.00) -- (6.12,0.00);
 \draw [thick, draw=black] (6.75,0.25) -- (6.02,0.25);
 \draw [thick, draw=black] (6.85, 0.50) -- (5.95,0.50);
 \node at (6.38,0.7){\scriptsize$a_{1},\, a_{1}^{\dagger}$};


 \draw [ultra thick,dashed,draw=black, fill=orange, fill opacity=0.2] (8.8,1.25) ellipse (0.65cm and 0.7cm);
 \draw [very thick, draw=black] (8.20,0.92) -- (9.20,0.92);
 \draw [very thick, draw=black] (8.20,1.60) -- (9.20,1.60);
 \node at (8.8, 1.76){\scriptsize $\ket{e}$};
 \node at (8.8, 0.75){\scriptsize $\ket{g}$};

 \draw [<->, dashed, thick, draw=black] (8.8,0.92) -- (8.8,1.6) node[pos=0.5,left] {$\omega$} ;

 \draw [ultra thick,dashed,draw=black, fill=orange, fill opacity=0.2] (8.8,-0.75) ellipse (0.65cm and 0.7cm);
 \draw [very thick, draw=black] (8.3,-1.1) -- (9.30,-1.1) ;
 \draw [very thick, draw=black] (8.3,-0.45) -- (9.30,-0.45) ;
 \node at (8.8, -0.25){\scriptsize $\ket{e}$};
 \node at (8.8, -1.25){\scriptsize $\ket{g}$};

\draw [<->, dashed, thick, draw=black] (8.8,-1.1) -- (8.8,-0.45) node[pos=0.5,left] {$\omega$} ;

 \draw [ultra thick,dashed,draw=black, fill=yellow, fill opacity=0.2] (11.2,0.25) ellipse (0.65cm and 0.7cm);
 \draw [very thick, draw=black] (10.72,0.65) parabola[parabola height=-1.0cm] (11.67,0.65);
 \draw [thick, draw=black] (11.05,-0.25) -- (11.35,-0.25);
 \draw [thick, draw=black] (10.92, 0.00) -- (11.47,0.00);
 \draw [thick, draw=black] (10.82, 0.25) -- (11.55,0.25);
 \draw [thick, draw=black] (10.74, 0.50) -- (11.65,0.50);
 \node at (11.16,0.7){\scriptsize$a_{2},\, a_{2}^{\dagger}$};

 \draw [<->, very thick, draw=black] (7.05, 0.50) -- (8.15,1.1);
 \draw [<->, very thick, draw=black] (7.05,-0.00) -- (8.10,-0.7);
 \draw [<->, very thick, draw=black] (9.47,1.10) -- (10.56,0.5);
 \draw [<->, very thick, draw=black] (9.49,-0.7) -- (10.56,-0.0);

 \node at (7.52,0.98) {\footnotesize JC};
 \node at (7.90,0.6) {\footnotesize $\kappa_{\text{JC}_{a1}}$};

 \node at (10.20,0.95) {\footnotesize JC};
 \node at (10.00,0.6) {\footnotesize $\kappa_{\text{JC}_{b1}}$};

 \node at (7.75,-0.20) {\footnotesize JC};
 \node at (7.50,-0.57) {\footnotesize $\kappa_{\text{JC}_{a2}}$};

 \node at (9.90,-0.2) {\footnotesize JC};
 \node at (10.35,-0.55) {\footnotesize $\kappa_{\text{JC}_{b2}}$};
 
\end{tikzpicture} 
\includegraphics{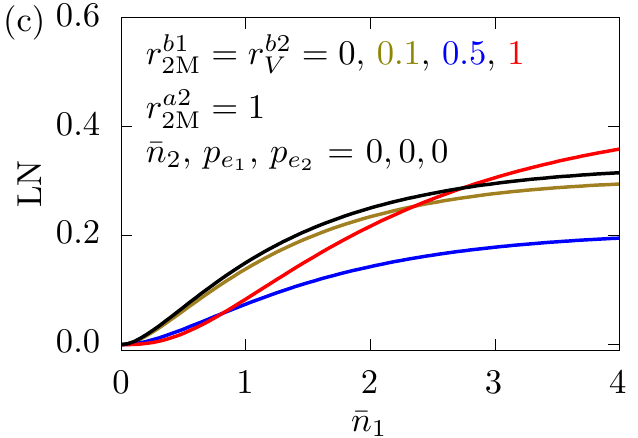} \\ \vspace{-0ex}
\caption{Schematics of the two configurations of the multiqubit systems represented by Hamiltonians $H_{\mathrm{2S}} $ (top) and $H_{\mathrm{2M}} $ (bottom left). Either both qubits are directly coupled to one of the oscillator pair, or both qubits together mediate the interaction between the oscillator pair. Each curve is calculated for an optimal time as described in the text. $r^{\text{BS}}_{\mathrm{2S}}$ and $r^{a2}_{\mathrm{2M}}$ are fixed at unity while $r_{\mathrm{2S}}^{b}$ and $r_{\mathrm{2M}}^{bi}$ are varied. For simplicity, we maintain the symmetry of $r_{\mathrm{2M}}^{b1} = r_{\mathrm{2M}}^{b2}$. (a-b) The results show the effect of increasing the thermal energy in one or other of the oscillators on the entanglement generation for $H_{\mathrm{2S}} $. (c) The symmetry of system $\mathrm{2M}$ means that the role of $\bar{n}_1$ and $\bar{n}_2$ can be interchanged for  specific ratio parameters. The black curves denote the degenerate cases when the systems $\mathrm{2S}$ and $\mathrm{2M}$ recapitulate systems $\mathrm{1S}$ and $\mathrm{1M}$.}
 \label{fig3}
\end{figure*}
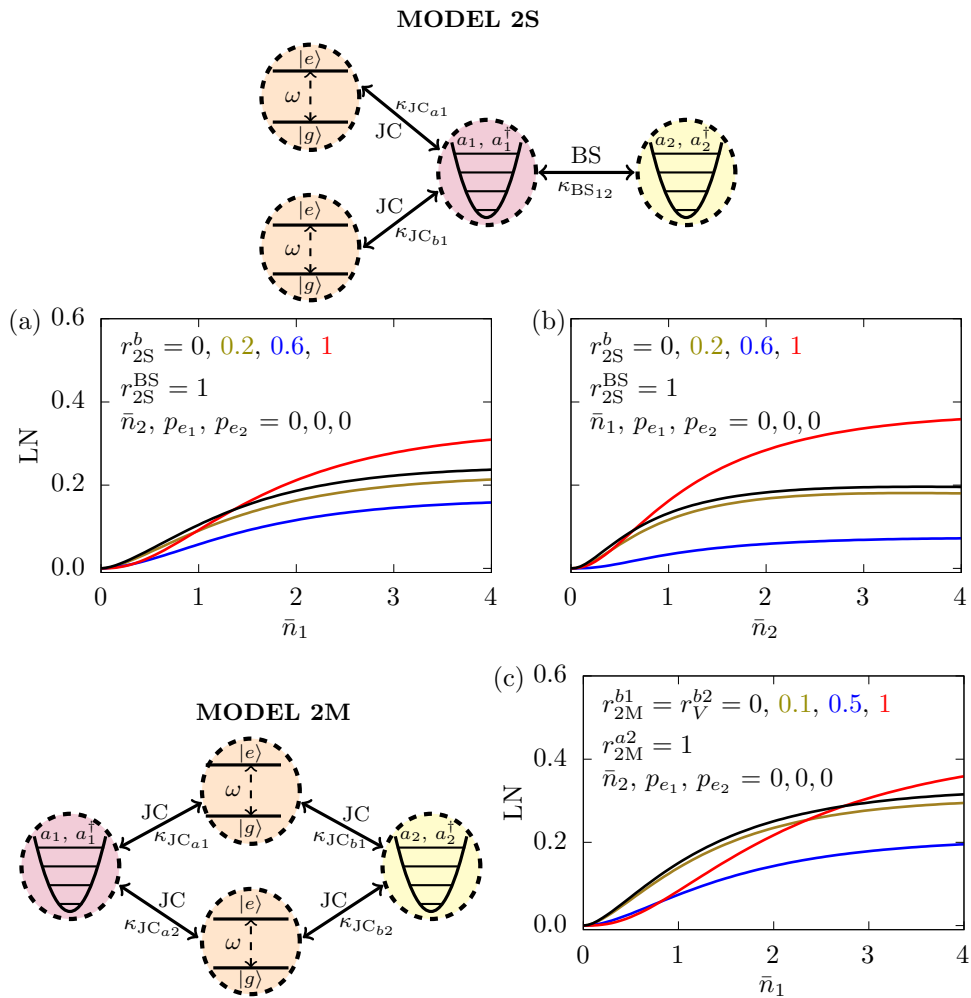

To distinguish qubit labels from oscillator labels we will use the lower-case Latin alphabet to label the qubits. In analysing the various configurations of qubits and oscillators we introduce a few more ratio parameters by defining $r_{\mathrm{2S}}^\text{BS}=\frac{\kappa_{\text{BS}_{12}}}{\kappa_{\text{JC}_{a1}}}$ and $r_{\mathrm{2S}}^b=\frac{\kappa_{\text{JC}_{b1}}}{\kappa_{\text{JC}_{a1}}}$ for system $\mathrm{2S}$, and $r_{\mathrm{2M}}^{bi}=\frac{\kappa_{\text{JC}_{bi}}}{\kappa_{\text{JC}_{a1}}}$ and $r_{\mathrm{2M}}^{a2} = \frac{\kappa_{\text{JC}_{a2}}}{\kappa_{\text{JC}_{a1}}}$ for system $\mathrm{2M}$ (Fig.~\ref{fig3}). Again, we study the dynamics in units of the scaled time $\tau = \kappa_{\text{JC}_{a1}}t$. In Fig.~\ref{fig3} we show a schematic of Hamiltonians $\mathrm{2S}$ and $\mathrm{2M}$, defined by
\begin{align}
\label{eqn:ham3}
H_{\mathrm{2S}} &= H_{\text{JC}_{a1}} + H_{\text{JC}_{b1}} + H_{\text{BS}_{12}}\,, \\
H_{\mathrm{2M}} &= H_{\text{JC}_{a1}} + H_{\text{JC}_{a2}} + H_{\text{JC}_{b1}} + H_{\text{JC}_{b2}}\,.
\label{eqn:ham_5}
\end{align}
$H_{\mathrm{2S}} $ is symmetric in the qubit labels and reduces to $H_{\mathrm{1S}} $ when either $\kappa_{\text{JC}_{a1}} = 0$ or $\kappa_{\text{JC}_{b1}} = 0$, while $H_{\mathrm{2M}} $, due to the qubit coupling symmetry, reduces to that of $H_{\mathrm{1M}} $ if either $\kappa_{\text{JC}_{a1}}=\kappa_{\text{JC}_{a2}}=0$ or $\kappa_{\text{JC}_{b1}}=\kappa_{\text{JC}_{b2}}=0$.

In Fig.~\ref{fig3} the black curves indicate the cases in which these systems reduce to their relevant single qubit cases. Therefore, any parameter regime which surpasses the black curve indicates an enhancement over the single-qubit case. For system $\mathrm{2S}$, panels (a) and (b) indicate that enhancement occurs for the symmetric coupling of the qubits to the oscillator (red curve). In the previous section we noted the importance of coupling symmetry, alongside nonlinearity and temperature difference, for inducing thermal entanglement. This is strongly reflected here. However, we also see a striking difference from system $\mathrm{1S}$ in that the significance of the temperature difference is inverted, so that the thermally excited distant oscillator is more significant for entanglement generation. Furthermore, crossing of the entanglement is visible and shows that there must be a sufficiently high temperature difference in order to gain enhancement. Panel (c) also demonstrates enhancement of entanglement for system $\mathrm{2M}$. As expected, the symmetry of the system means that the behaviour with respect to $\bar{n}_1$ and $\bar{n}_2$ is identical. The same requirement of symmetry occurs in order to achieve an enhanced generation of entanglement, and the same crossing occurs but with the requirement that the temperature difference is higher.   

Finally we remark that a third configuration is given by $H_{\mathrm{1S-1S}}$, where each qubit separately interacts with each oscillator. We write 
\begin{align}
H_{\mathrm{1S-1S}}  = H_{\text{JC}_{a1}}+H_{\text{JC}_{b2}}+H_{\text{BS}_{12}}\,.
\label{eqn:ham4}
\end{align}
Similar to $H_{\mathrm{2S}} $, $H_{\mathrm{1S-1S}}$ also reduces to $H_{\mathrm{1S}} $ when either $\kappa_{\text{JC}_{a1}} = 0$ or $\kappa_{\text{JC}_{b1}} = 0$. Our analysis reveals that $H_{\mathrm{1S-1S}}$ fails to produce any enhancement of entanglement than what is already present in $H_{\mathrm{1S}} $ in most parameter regimes.

\section{\label{sec:discussions} Discussion}

We have demonstrated the autonomous generation of entanglement in a pair of linearly interacting oscillators, enhanced by the temperature difference of the oscillators and driven by the anharmonic modulation of their motion via discrete variable (DV) systems. This effect is of a different nature than  previous examples of thermally driven entanglement generation between DV systems, driven either by discrete or continuous variable (CV) systems. The entangling of the oscillators is a nontrivial outworking of the nonlinear system as evidenced by the emergence of large-dimensional entanglement only after a significant interaction time. While the temperature difference drives the entanglement generation, this eventually saturates. A two-qubit maximally entangled state has $\text{LN} = 1$ and a pair of continuous variable systems are capable of storing much more than that~\cite{kim_accumulation_2007}, however these require additional resources beyond thermal energy. With the oscillators initially in their respective ground states, a population inverted qubit ($p_e=1$) indeed maximally generates 1 ebit in the oscillators, suggesting that the entanglement of the system is saturated by the capacity of the qubit. With a thermal qubit ($0\le p_e\le\frac12$) and ground state oscillators however this is limited to $\text{LN} \approx 0.27$, which is much lower. If the thermal resources are associated with the oscillators however, we find in at least one case (see Fig.~\ref{fig2}~(a)) that $\text{LN} = 0.5$ which is almost twice that available from a thermally excited qubit. That said, thermal resources can coherently push beyond a thermal equilibrium state~\cite{cirac_pra_1994}. A thermal oscillator interacting with a ground state qubit via the JC coupling can induce partial population inversion with $p_e\approx 0.63$. A qubit prepared in such a way, beyond thermal equilibrium, can subsequently be coupled to a ground state oscillator pair and induce entanglement with $\text{LN}\approx 0.44$, larger than $\text{LN}\approx 0.27$ achievable for incoherent thermal equilibrium. Whereas $\text{LN}\approx 0.27$ can be overcome by both $\mathrm{1S}$ and $\mathrm{1M}$, $\text{LN}=0.44$ is overcome only by $\mathrm{1S}$ (see the Supplementary Material for a detailed analysis). This demonstrates the superiority of a scheme based on the thermal states of the oscillators and a CV beamsplitter class of dynamics.

Since thermal resources are exhausted, what remains is that the nonlinear resource of the qubit is over capacity. This is supported by the enhancement effects of multiqubit systems, which only take effect once the thermal noise is sufficiently high. However, other nonlinear interactions with oscillators may have even stronger potential for thermally induced nonlinearities. We note in general that symmetry in interactions, where available, supports the generation of entanglement. This contrasts with the asymmetry required in the thermal energy of the oscillators. Our demonstration of thermally induced entanglement, spread over a large dimension in the two-oscillator Hilbert space has remained unnoticed for a long time, despite being a natural step from previous analyses. A reason might be that the natural method in platforms such as trapped ions is to dynamically transfer the entanglement to the qubits where readout methods are more developed~\cite{myerson_high-fidelity_2008}. We investigated the possibility of using a general qubit-oscillator interaction, to transfer entanglement from the oscillator pair to a qubit pair. This appears to fail, consistent with older results~\cite{paternostro_entanglement_2006}, and suggests that the large-dimensional entanglement induced by the qubits is curiously incompatible with qubit readout.

The coherence of the qubit interaction with the oscillator is absolutely necessary for entanglement generation, which we have verified in the Supplementary Material by analysing decoherence effects, particularly dephasing of the qubit. Large qubit dephasing is detrimental for entanglement generation. For small dephasing, however, the behaviour is qualitatively the same but with the LN reduced. In general the decoherence rates for trapped ion systems are below the timescales required to generate entanglement, so that while quantum features are typically sensitive to such processes they are expected to have negligible effects in our scheme. 

Using thermal states to induce entanglement indicates that an equilibrium probability distribution over the Fock basis has the possibility to induce quantum effects in linear oscillators when coupled to an ancillary ground-state nonlinear system. While inducing entanglement by thermal states is the main result of this article, we have also investigated coherent states and phase randomized coherent states. Both of them will not produce any entanglement by a linear passive coupling in the oscillator, but they are less noisy than thermal states. Taking coherent states as examples, we find that the generation of entanglement is greatly enhanced when using such resources as demonstrated in the Supplementary Material. Similarly, phase randomised coherent states (PRCS), which implement a classical Poisson distribution over the Fock states rather than a Boltzmann distribution, also produce larger entanglement than thermal states although still less than coherent states. Both the PRCS and the thermal states are mixtures over the Fock basis and cannot produce entanglement with only linear coupling. For the same mean phonon number, Poissonian statistics perform significantly better than Boltzmann statistics due to less noise in the phonon number. As mentioned, if autonomous quantum systems are to be achieved both entanglement and the coherent states must be available unconditionally and autonomously. If quantum coherence can enhance entanglement generation as we have seen with coherent states, then thermally induced quantum coherence, if possible, will be a particularly powerful tool for autonomous quantum systems. However, that is a separate problem for the oscillator equivalent to the entire problem of shot-noise-limited lasing without an external drive.

The enhancement of the entanglement generation by complex DV systems bears some relation to existing generalisations of the JC interaction such as the Tavis-Cummings model or the Dicke model. Increasing the number of two-level systems is one method to address increasing the capacity of the nonlinearity. Another is to increase the dimension of the DV system by addressing more than two energy levels in the motion induced by the dipole interaction. The scaling of the entanglement dynamics with multilevel subsystems may unveil the details of how hybrid quantum systems eventually build deterministic nonlinearities for the oscillator. Adding more ions also adds more sets of passively coupled oscillators, allowing for the possibility to autonomously generate multipartite entanglement beyond Gaussian approximations~\cite{adesso_genuine_2008}. This possibility is a natural further extension of our investigation and bears strongly on the potential existence of new autonomous quantum effects in many-body systems. Alternatively, the procedure developed here can be repeated multiple times with a newly prepared ground state qubit in order to accumulate entanglement. Such a technique has already been successfully used to accumulate thermally induced nonclassical features of a single oscillator~\cite{podhora_unconditional_2020}.

When an experiment becomes available to demonstrate these phenomena, it will stimulate further development of thermally induced entanglement of oscillators on a larger scale. In support of that end, a small quantum simulator of such effects will unlock a deeper understanding of these phenomena and the techniques required to certify their feasibility. Any form of entanglement is, by definition, conclusive proof that two systems interacted in the past in a genuinely quantum mechanical way, outside the class of local operations and classical communication (LOCC)~\cite{peres}. The growth in entanglement from separable states then provides evidence that a more extensive exchange of quanta occurred than that which occurs between thermal linear oscillators. That is, a genuine quantum mechanical interaction must have taken place, such as coupling to two-level systems. This hysteresis is essential for developing platforms for quantum technology and can be simulated and investigated using a small-scale hybrid quantum simulator. The results of such an investigation will aid new platforms involving hybrid quantum physics~\cite{golter,gartner,arita,cyang,rogers_hybrid_2014,andersen_hybrid_2015,chu_quantum_2017,cady_diamond_2019} to navigate through the discoveries of quantum interactions between oscillators.

\acknowledgements  
The authors acknowledge funding from Project 21-13265X of the Czech Science Foundation. 

\bibliography{reference}

\begin{thebibliography}{64}%
\makeatletter
\providecommand \@ifxundefined [1]{%
 \@ifx{#1\undefined}
}%
\providecommand \@ifnum [1]{%
 \ifnum #1\expandafter \@firstoftwo
 \else \expandafter \@secondoftwo
 \fi
}%
\providecommand \@ifx [1]{%
 \ifx #1\expandafter \@firstoftwo
 \else \expandafter \@secondoftwo
 \fi
}%
\providecommand \natexlab [1]{#1}%
\providecommand \enquote  [1]{``#1''}%
\providecommand \bibnamefont  [1]{#1}%
\providecommand \bibfnamefont [1]{#1}%
\providecommand \citenamefont [1]{#1}%
\providecommand \href@noop [0]{\@secondoftwo}%
\providecommand \href [0]{\begingroup \@sanitize@url \@href}%
\providecommand \@href[1]{\@@startlink{#1}\@@href}%
\providecommand \@@href[1]{\endgroup#1\@@endlink}%
\providecommand \@sanitize@url [0]{\catcode `\\12\catcode `\$12\catcode
  `\&12\catcode `\#12\catcode `\^12\catcode `\_12\catcode `\%12\relax}%
\providecommand \@@startlink[1]{}%
\providecommand \@@endlink[0]{}%
\providecommand \url  [0]{\begingroup\@sanitize@url \@url }%
\providecommand \@url [1]{\endgroup\@href {#1}{\urlprefix }}%
\providecommand \urlprefix  [0]{URL }%
\providecommand \Eprint [0]{\href }%
\providecommand \doibase [0]{http://dx.doi.org/}%
\providecommand \selectlanguage [0]{\@gobble}%
\providecommand \bibinfo  [0]{\@secondoftwo}%
\providecommand \bibfield  [0]{\@secondoftwo}%
\providecommand \translation [1]{[#1]}%
\providecommand \BibitemOpen [0]{}%
\providecommand \bibitemStop [0]{}%
\providecommand \bibitemNoStop [0]{.\EOS\space}%
\providecommand \EOS [0]{\spacefactor3000\relax}%
\providecommand \BibitemShut  [1]{\csname bibitem#1\endcsname}%
\let\auto@bib@innerbib\@empty
\bibitem [{\citenamefont {Brask}\ \emph {et~al.}(2015)\citenamefont {Brask},
  \citenamefont {Haack}, \citenamefont {Brunner},\ and\ \citenamefont
  {Huber}}]{Bohr_Brask_2015}%
  \BibitemOpen
  \bibfield  {author} {\bibinfo {author} {\bibfnamefont {J.~B.}\ \bibnamefont
  {Brask}}, \bibinfo {author} {\bibfnamefont {G.}~\bibnamefont {Haack}},
  \bibinfo {author} {\bibfnamefont {N.}~\bibnamefont {Brunner}}, \ and\
  \bibinfo {author} {\bibfnamefont {M.}~\bibnamefont {Huber}},\ }\href
  {\doibase 10.1088/1367-2630/17/11/113029} {\bibfield  {journal} {\bibinfo
  {journal} {New J. Phys.}\ }\textbf {\bibinfo {volume} {17}},\ \bibinfo
  {pages} {113029} (\bibinfo {year} {2015})}\BibitemShut {NoStop}%
\bibitem [{\citenamefont {Bose}\ \emph {et~al.}(2001)\citenamefont {Bose},
  \citenamefont {Fuentes-Guridi}, \citenamefont {Knight},\ and\ \citenamefont
  {Vedral}}]{bose}%
  \BibitemOpen
  \bibfield  {author} {\bibinfo {author} {\bibfnamefont {S.}~\bibnamefont
  {Bose}}, \bibinfo {author} {\bibfnamefont {I.}~\bibnamefont
  {Fuentes-Guridi}}, \bibinfo {author} {\bibfnamefont {P.~L.}\ \bibnamefont
  {Knight}}, \ and\ \bibinfo {author} {\bibfnamefont {V.}~\bibnamefont
  {Vedral}},\ }\href {\doibase 10.1103/PhysRevLett.87.050401} {\bibfield
  {journal} {\bibinfo  {journal} {Phys. Rev. Lett.}\ }\textbf {\bibinfo
  {volume} {87}},\ \bibinfo {pages} {050401} (\bibinfo {year}
  {2001})}\BibitemShut {NoStop}%
\bibitem [{\citenamefont {Kim}\ \emph {et~al.}(2002)\citenamefont {Kim},
  \citenamefont {Lee}, \citenamefont {Ahn},\ and\ \citenamefont
  {Knight}}]{mskim}%
  \BibitemOpen
  \bibfield  {author} {\bibinfo {author} {\bibfnamefont {M.~S.}\ \bibnamefont
  {Kim}}, \bibinfo {author} {\bibfnamefont {J.}~\bibnamefont {Lee}}, \bibinfo
  {author} {\bibfnamefont {D.}~\bibnamefont {Ahn}}, \ and\ \bibinfo {author}
  {\bibfnamefont {P.~L.}\ \bibnamefont {Knight}},\ }\href {\doibase
  10.1103/PhysRevA.65.040101} {\bibfield  {journal} {\bibinfo  {journal} {Phys.
  Rev. A}\ }\textbf {\bibinfo {volume} {65}},\ \bibinfo {pages} {040101}
  (\bibinfo {year} {2002})}\BibitemShut {NoStop}%
\bibitem [{\citenamefont {Benatti}\ and\ \citenamefont
  {Floreanini}(2006)}]{benatti_entangling_2006}%
  \BibitemOpen
  \bibfield  {author} {\bibinfo {author} {\bibfnamefont {F.}~\bibnamefont
  {Benatti}}\ and\ \bibinfo {author} {\bibfnamefont {R.}~\bibnamefont
  {Floreanini}},\ }\href {\doibase 10.1088/0305-4470/39/11/009} {\bibfield
  {journal} {\bibinfo  {journal} {J. Phys. A: Math. Gen.}\ }\textbf {\bibinfo
  {volume} {39}},\ \bibinfo {pages} {2689} (\bibinfo {year}
  {2006})}\BibitemShut {NoStop}%
\bibitem [{\citenamefont {Zhou}\ \emph {et~al.}(2002)\citenamefont {Zhou},
  \citenamefont {Song},\ and\ \citenamefont {Li}}]{zhou_entanglement_2002}%
  \BibitemOpen
  \bibfield  {author} {\bibinfo {author} {\bibfnamefont {L.}~\bibnamefont
  {Zhou}}, \bibinfo {author} {\bibfnamefont {H.~S.}\ \bibnamefont {Song}}, \
  and\ \bibinfo {author} {\bibfnamefont {C.}~\bibnamefont {Li}},\ }\href
  {\doibase 10.1088/1464-4266/4/6/310} {\bibfield  {journal} {\bibinfo
  {journal} {J. Opt. B: Quantum Semiclass. Opt.}\ }\textbf {\bibinfo {volume}
  {4}},\ \bibinfo {pages} {425} (\bibinfo {year} {2002})}\BibitemShut {NoStop}%
\bibitem [{\citenamefont {Brádler}\ and\ \citenamefont
  {Jáuregui}(2007)}]{bradler_entanglement_2007}%
  \BibitemOpen
  \bibfield  {author} {\bibinfo {author} {\bibfnamefont {K.}~\bibnamefont
  {Brádler}}\ and\ \bibinfo {author} {\bibfnamefont {R.}~\bibnamefont
  {Jáuregui}},\ }\href {\doibase 10.1088/0953-4075/40/4/009} {\bibfield
  {journal} {\bibinfo  {journal} {J. Phys. B: At. Mol. Opt. Phys}\ }\textbf
  {\bibinfo {volume} {40}},\ \bibinfo {pages} {743} (\bibinfo {year}
  {2007})}\BibitemShut {NoStop}%
\bibitem [{\citenamefont {Ping}\ \emph {et~al.}(2008)\citenamefont {Ping},
  \citenamefont {Bin}, \citenamefont {Zhao-Ming},\ and\ \citenamefont
  {Jian}}]{ping_continuous_2008}%
  \BibitemOpen
  \bibfield  {author} {\bibinfo {author} {\bibfnamefont {C.}~\bibnamefont
  {Ping}}, \bibinfo {author} {\bibfnamefont {S.}~\bibnamefont {Bin}}, \bibinfo
  {author} {\bibfnamefont {W.}~\bibnamefont {Zhao-Ming}}, \ and\ \bibinfo
  {author} {\bibfnamefont {Z.}~\bibnamefont {Jian}},\ }\href {\doibase
  10.1088/0253-6102/49/6/59} {\bibfield  {journal} {\bibinfo  {journal}
  {Commun. Theor. Phys.}\ }\textbf {\bibinfo {volume} {49}},\ \bibinfo {pages}
  {1622} (\bibinfo {year} {2008})}\BibitemShut {NoStop}%
\bibitem [{\citenamefont {Patrick}\ \emph {et~al.}(2011)\citenamefont
  {Patrick}, \citenamefont {Yang}, \citenamefont {Yin},\ and\ \citenamefont
  {Li}}]{patrick_entangling_2011}%
  \BibitemOpen
  \bibfield  {author} {\bibinfo {author} {\bibfnamefont {S.~R.~J.}\
  \bibnamefont {Patrick}}, \bibinfo {author} {\bibfnamefont {Y.}~\bibnamefont
  {Yang}}, \bibinfo {author} {\bibfnamefont {Z.-Q.}\ \bibnamefont {Yin}}, \
  and\ \bibinfo {author} {\bibfnamefont {F.-L.}\ \bibnamefont {Li}},\ }\href
  {\doibase 10.1142/S0217979211101478} {\bibfield  {journal} {\bibinfo
  {journal} {Int. J. Mod. Phys. B}\ }\textbf {\bibinfo {volume} {25}},\
  \bibinfo {pages} {2681} (\bibinfo {year} {2011})}\BibitemShut {NoStop}%
\bibitem [{\citenamefont {Filip}\ \emph {et~al.}(2002)\citenamefont {Filip},
  \citenamefont {Du\ifmmode~\check{s}\else \v{s}\fi{}ek}, \citenamefont
  {Fiur\'a\ifmmode~\check{s}\else \v{s}\fi{}ek},\ and\ \citenamefont
  {Mi\ifmmode~\check{s}\else \v{s}\fi{}ta}}]{filip02}%
  \BibitemOpen
  \bibfield  {author} {\bibinfo {author} {\bibfnamefont {R.}~\bibnamefont
  {Filip}}, \bibinfo {author} {\bibfnamefont {M.}~\bibnamefont
  {Du\ifmmode~\check{s}\else \v{s}\fi{}ek}}, \bibinfo {author} {\bibfnamefont
  {J.}~\bibnamefont {Fiur\'a\ifmmode~\check{s}\else \v{s}\fi{}ek}}, \ and\
  \bibinfo {author} {\bibfnamefont {L.}~\bibnamefont {Mi\ifmmode~\check{s}\else
  \v{s}\fi{}ta}},\ }\href {\doibase 10.1103/PhysRevA.65.043802} {\bibfield
  {journal} {\bibinfo  {journal} {Phys. Rev. A}\ }\textbf {\bibinfo {volume}
  {65}},\ \bibinfo {pages} {043802} (\bibinfo {year} {2002})}\BibitemShut
  {NoStop}%
\bibitem [{\citenamefont {Gao}\ \emph {et~al.}(2019)\citenamefont {Gao},
  \citenamefont {Lester}, \citenamefont {Chou}, \citenamefont {Frunzio},
  \citenamefont {Devoret}, \citenamefont {Jiang}, \citenamefont {Girvin},\ and\
  \citenamefont {Schoelkopf}}]{yygao}%
  \BibitemOpen
  \bibfield  {author} {\bibinfo {author} {\bibfnamefont {Y.~Y.}\ \bibnamefont
  {Gao}}, \bibinfo {author} {\bibfnamefont {B.~J.}\ \bibnamefont {Lester}},
  \bibinfo {author} {\bibfnamefont {K.~S.}\ \bibnamefont {Chou}}, \bibinfo
  {author} {\bibfnamefont {L.}~\bibnamefont {Frunzio}}, \bibinfo {author}
  {\bibfnamefont {M.~H.}\ \bibnamefont {Devoret}}, \bibinfo {author}
  {\bibfnamefont {L.}~\bibnamefont {Jiang}}, \bibinfo {author} {\bibfnamefont
  {S.~M.}\ \bibnamefont {Girvin}}, \ and\ \bibinfo {author} {\bibfnamefont
  {R.~J.}\ \bibnamefont {Schoelkopf}},\ }\href {\doibase
  10.1038/s41586-019-0970-4} {\bibfield  {journal} {\bibinfo  {journal}
  {Nature}\ }\textbf {\bibinfo {volume} {566}},\ \bibinfo {pages} {509}
  (\bibinfo {year} {2019})}\BibitemShut {NoStop}%
\bibitem [{\citenamefont {Gan}\ \emph {et~al.}(2020)\citenamefont {Gan},
  \citenamefont {Maslennikov}, \citenamefont {Tseng}, \citenamefont {Nguyen},\
  and\ \citenamefont {Matsukevich}}]{hgan}%
  \BibitemOpen
  \bibfield  {author} {\bibinfo {author} {\bibfnamefont {H.~C.~J.}\
  \bibnamefont {Gan}}, \bibinfo {author} {\bibfnamefont {G.}~\bibnamefont
  {Maslennikov}}, \bibinfo {author} {\bibfnamefont {K.-W.}\ \bibnamefont
  {Tseng}}, \bibinfo {author} {\bibfnamefont {C.}~\bibnamefont {Nguyen}}, \
  and\ \bibinfo {author} {\bibfnamefont {D.}~\bibnamefont {Matsukevich}},\
  }\href {\doibase 10.1103/PhysRevLett.124.170502} {\bibfield  {journal}
  {\bibinfo  {journal} {Phys. Rev. Lett.}\ }\textbf {\bibinfo {volume} {124}},\
  \bibinfo {pages} {170502} (\bibinfo {year} {2020})}\BibitemShut {NoStop}%
\bibitem [{\citenamefont {Filip}\ and\ \citenamefont {Marek}(2014)}]{filip14}%
  \BibitemOpen
  \bibfield  {author} {\bibinfo {author} {\bibfnamefont {R.}~\bibnamefont
  {Filip}}\ and\ \bibinfo {author} {\bibfnamefont {P.}~\bibnamefont {Marek}},\
  }\href {\doibase 10.1103/PhysRevA.90.063820} {\bibfield  {journal} {\bibinfo
  {journal} {Phys. Rev. A}\ }\textbf {\bibinfo {volume} {90}},\ \bibinfo
  {pages} {063820} (\bibinfo {year} {2014})}\BibitemShut {NoStop}%
\bibitem [{\citenamefont {Slodi\v{c}ka}\ \emph {et~al.}(2016)\citenamefont
  {Slodi\v{c}ka}, \citenamefont {Marek},\ and\ \citenamefont
  {Filip}}]{slodicka}%
  \BibitemOpen
  \bibfield  {author} {\bibinfo {author} {\bibfnamefont {L.}~\bibnamefont
  {Slodi\v{c}ka}}, \bibinfo {author} {\bibfnamefont {P.}~\bibnamefont {Marek}},
  \ and\ \bibinfo {author} {\bibfnamefont {R.}~\bibnamefont {Filip}},\ }\href
  {\doibase 10.1364/OE.24.007858} {\bibfield  {journal} {\bibinfo  {journal}
  {Opt. Express}\ }\textbf {\bibinfo {volume} {24}},\ \bibinfo {pages} {7858}
  (\bibinfo {year} {2016})}\BibitemShut {NoStop}%
\bibitem [{\citenamefont {Marek}\ \emph {et~al.}(2016)\citenamefont {Marek},
  \citenamefont {Lachman}, \citenamefont {Slodi\v{c}ka},\ and\ \citenamefont
  {Filip}}]{marek}%
  \BibitemOpen
  \bibfield  {author} {\bibinfo {author} {\bibfnamefont {P.}~\bibnamefont
  {Marek}}, \bibinfo {author} {\bibfnamefont {L.}~\bibnamefont {Lachman}},
  \bibinfo {author} {\bibfnamefont {L.}~\bibnamefont {Slodi\v{c}ka}}, \ and\
  \bibinfo {author} {\bibfnamefont {R.}~\bibnamefont {Filip}},\ }\href
  {\doibase 10.1103/PhysRevA.94.013850} {\bibfield  {journal} {\bibinfo
  {journal} {Phys. Rev. A}\ }\textbf {\bibinfo {volume} {94}},\ \bibinfo
  {pages} {013850} (\bibinfo {year} {2016})}\BibitemShut {NoStop}%
\bibitem [{\citenamefont {Kienzler}\ \emph {et~al.}(2015)\citenamefont
  {Kienzler}, \citenamefont {Lo}, \citenamefont {Keitch}, \citenamefont
  {de~Clercq}, \citenamefont {Leupold}, \citenamefont {Lindenfelser},
  \citenamefont {Marinelli}, \citenamefont {Negnevitsky},\ and\ \citenamefont
  {Home}}]{kienzler_quantum_2015}%
  \BibitemOpen
  \bibfield  {author} {\bibinfo {author} {\bibfnamefont {D.}~\bibnamefont
  {Kienzler}}, \bibinfo {author} {\bibfnamefont {H.-Y.}\ \bibnamefont {Lo}},
  \bibinfo {author} {\bibfnamefont {B.}~\bibnamefont {Keitch}}, \bibinfo
  {author} {\bibfnamefont {L.}~\bibnamefont {de~Clercq}}, \bibinfo {author}
  {\bibfnamefont {F.}~\bibnamefont {Leupold}}, \bibinfo {author} {\bibfnamefont
  {F.}~\bibnamefont {Lindenfelser}}, \bibinfo {author} {\bibfnamefont
  {M.}~\bibnamefont {Marinelli}}, \bibinfo {author} {\bibfnamefont
  {V.}~\bibnamefont {Negnevitsky}}, \ and\ \bibinfo {author} {\bibfnamefont
  {J.~P.}\ \bibnamefont {Home}},\ }\href {\doibase 10.1126/science.1261033}
  {\bibfield  {journal} {\bibinfo  {journal} {Science}\ }\textbf {\bibinfo
  {volume} {347}},\ \bibinfo {pages} {53} (\bibinfo {year} {2015})}\BibitemShut
  {NoStop}%
\bibitem [{\citenamefont {Johanning}\ \emph {et~al.}(2009)\citenamefont
  {Johanning}, \citenamefont {Var{\'{o}}n},\ and\ \citenamefont
  {Wunderlich}}]{johanning}%
  \BibitemOpen
  \bibfield  {author} {\bibinfo {author} {\bibfnamefont {M.}~\bibnamefont
  {Johanning}}, \bibinfo {author} {\bibfnamefont {A.~F.}\ \bibnamefont
  {Var{\'{o}}n}}, \ and\ \bibinfo {author} {\bibfnamefont {C.}~\bibnamefont
  {Wunderlich}},\ }\href {\doibase 10.1088/0953-4075/42/15/154009} {\bibfield
  {journal} {\bibinfo  {journal} {J. Phys. B: At. Mol. Opt. Phys}\ }\textbf
  {\bibinfo {volume} {42}},\ \bibinfo {pages} {154009} (\bibinfo {year}
  {2009})}\BibitemShut {NoStop}%
\bibitem [{\citenamefont {Blatt}\ and\ \citenamefont {Roos}(2012)}]{blatt}%
  \BibitemOpen
  \bibfield  {author} {\bibinfo {author} {\bibfnamefont {R.}~\bibnamefont
  {Blatt}}\ and\ \bibinfo {author} {\bibfnamefont {C.~F.}\ \bibnamefont
  {Roos}},\ }\href {\doibase 10.1038/nphys2252} {\bibfield  {journal} {\bibinfo
   {journal} {Nat. Phys.}\ }\textbf {\bibinfo {volume} {8}},\ \bibinfo {pages}
  {277} (\bibinfo {year} {2012})}\BibitemShut {NoStop}%
\bibitem [{\citenamefont {Friedenauer}\ \emph {et~al.}(2008)\citenamefont
  {Friedenauer}, \citenamefont {Schmitz}, \citenamefont {Glueckert},
  \citenamefont {Porras},\ and\ \citenamefont {Schaetz}}]{friedenauer}%
  \BibitemOpen
  \bibfield  {author} {\bibinfo {author} {\bibfnamefont {A.}~\bibnamefont
  {Friedenauer}}, \bibinfo {author} {\bibfnamefont {H.}~\bibnamefont
  {Schmitz}}, \bibinfo {author} {\bibfnamefont {J.~T.}\ \bibnamefont
  {Glueckert}}, \bibinfo {author} {\bibfnamefont {D.}~\bibnamefont {Porras}}, \
  and\ \bibinfo {author} {\bibfnamefont {T.}~\bibnamefont {Schaetz}},\ }\href
  {\doibase 10.1038/nphys1032} {\bibfield  {journal} {\bibinfo  {journal} {Nat.
  Phys.}\ }\textbf {\bibinfo {volume} {4}},\ \bibinfo {pages} {757} (\bibinfo
  {year} {2008})}\BibitemShut {NoStop}%
\bibitem [{\citenamefont {Kim}\ \emph {et~al.}(2010)\citenamefont {Kim},
  \citenamefont {Chang}, \citenamefont {Korenblit}, \citenamefont {Islam},
  \citenamefont {Edwards}, \citenamefont {Freericks}, \citenamefont {Lin},
  \citenamefont {Duan},\ and\ \citenamefont {Monroe}}]{kim1}%
  \BibitemOpen
  \bibfield  {author} {\bibinfo {author} {\bibfnamefont {K.}~\bibnamefont
  {Kim}}, \bibinfo {author} {\bibfnamefont {M.~S.}\ \bibnamefont {Chang}},
  \bibinfo {author} {\bibfnamefont {S.}~\bibnamefont {Korenblit}}, \bibinfo
  {author} {\bibfnamefont {R.}~\bibnamefont {Islam}}, \bibinfo {author}
  {\bibfnamefont {E.~E.}\ \bibnamefont {Edwards}}, \bibinfo {author}
  {\bibfnamefont {J.~K.}\ \bibnamefont {Freericks}}, \bibinfo {author}
  {\bibfnamefont {G.~D.}\ \bibnamefont {Lin}}, \bibinfo {author} {\bibfnamefont
  {L.~M.}\ \bibnamefont {Duan}}, \ and\ \bibinfo {author} {\bibfnamefont
  {C.}~\bibnamefont {Monroe}},\ }\href {\doibase 10.1038/nature09071}
  {\bibfield  {journal} {\bibinfo  {journal} {Nature}\ }\textbf {\bibinfo
  {volume} {465}},\ \bibinfo {pages} {590} (\bibinfo {year}
  {2010})}\BibitemShut {NoStop}%
\bibitem [{\citenamefont {Kim}\ \emph {et~al.}(2009)\citenamefont {Kim},
  \citenamefont {Chang}, \citenamefont {Islam}, \citenamefont {Korenblit},
  \citenamefont {Duan},\ and\ \citenamefont {Monroe}}]{kim2}%
  \BibitemOpen
  \bibfield  {author} {\bibinfo {author} {\bibfnamefont {K.}~\bibnamefont
  {Kim}}, \bibinfo {author} {\bibfnamefont {M.-S.}\ \bibnamefont {Chang}},
  \bibinfo {author} {\bibfnamefont {R.}~\bibnamefont {Islam}}, \bibinfo
  {author} {\bibfnamefont {S.}~\bibnamefont {Korenblit}}, \bibinfo {author}
  {\bibfnamefont {L.-M.}\ \bibnamefont {Duan}}, \ and\ \bibinfo {author}
  {\bibfnamefont {C.}~\bibnamefont {Monroe}},\ }\href {\doibase
  10.1103/PhysRevLett.103.120502} {\bibfield  {journal} {\bibinfo  {journal}
  {Phys. Rev. Lett.}\ }\textbf {\bibinfo {volume} {103}},\ \bibinfo {pages}
  {120502} (\bibinfo {year} {2009})}\BibitemShut {NoStop}%
\bibitem [{\citenamefont {Islam}\ \emph {et~al.}(2011)\citenamefont {Islam},
  \citenamefont {Edwards}, \citenamefont {Kim}, \citenamefont {Korenblit},
  \citenamefont {Noh}, \citenamefont {Carmichael}, \citenamefont {Lin},
  \citenamefont {Duan}, \citenamefont {Joseph~Wang}, \citenamefont
  {Freericks},\ and\ \citenamefont {Monroe}}]{islam}%
  \BibitemOpen
  \bibfield  {author} {\bibinfo {author} {\bibfnamefont {R.}~\bibnamefont
  {Islam}}, \bibinfo {author} {\bibfnamefont {E.~E.}\ \bibnamefont {Edwards}},
  \bibinfo {author} {\bibfnamefont {K.}~\bibnamefont {Kim}}, \bibinfo {author}
  {\bibfnamefont {S.}~\bibnamefont {Korenblit}}, \bibinfo {author}
  {\bibfnamefont {C.}~\bibnamefont {Noh}}, \bibinfo {author} {\bibfnamefont
  {H.}~\bibnamefont {Carmichael}}, \bibinfo {author} {\bibfnamefont {G.~D.}\
  \bibnamefont {Lin}}, \bibinfo {author} {\bibfnamefont {L.~M.}\ \bibnamefont
  {Duan}}, \bibinfo {author} {\bibfnamefont {C.~C.}\ \bibnamefont
  {Joseph~Wang}}, \bibinfo {author} {\bibfnamefont {J.~K.}\ \bibnamefont
  {Freericks}}, \ and\ \bibinfo {author} {\bibfnamefont {C.}~\bibnamefont
  {Monroe}},\ }\href {\doibase 10.1038/ncomms1374} {\bibfield  {journal}
  {\bibinfo  {journal} {Nat. Commun.}\ }\textbf {\bibinfo {volume} {2}},\
  \bibinfo {pages} {377} (\bibinfo {year} {2011})}\BibitemShut {NoStop}%
\bibitem [{\citenamefont {van~der Zant}\ \emph {et~al.}(1992)\citenamefont
  {van~der Zant}, \citenamefont {Fritschy}, \citenamefont {Elion},
  \citenamefont {Geerligs},\ and\ \citenamefont {Mooij}}]{zant}%
  \BibitemOpen
  \bibfield  {author} {\bibinfo {author} {\bibfnamefont {H.~S.~J.}\
  \bibnamefont {van~der Zant}}, \bibinfo {author} {\bibfnamefont {F.~C.}\
  \bibnamefont {Fritschy}}, \bibinfo {author} {\bibfnamefont {W.~J.}\
  \bibnamefont {Elion}}, \bibinfo {author} {\bibfnamefont {L.~J.}\ \bibnamefont
  {Geerligs}}, \ and\ \bibinfo {author} {\bibfnamefont {J.~E.}\ \bibnamefont
  {Mooij}},\ }\href {\doibase 10.1103/PhysRevLett.69.2971} {\bibfield
  {journal} {\bibinfo  {journal} {Phys. Rev. Lett.}\ }\textbf {\bibinfo
  {volume} {69}},\ \bibinfo {pages} {2971} (\bibinfo {year}
  {1992})}\BibitemShut {NoStop}%
\bibitem [{\citenamefont {Greiner}\ \emph {et~al.}(2002)\citenamefont
  {Greiner}, \citenamefont {Mandel}, \citenamefont {Esslinger}, \citenamefont
  {H\"{a}nsch},\ and\ \citenamefont {Bloch}}]{greiner}%
  \BibitemOpen
  \bibfield  {author} {\bibinfo {author} {\bibfnamefont {M.}~\bibnamefont
  {Greiner}}, \bibinfo {author} {\bibfnamefont {O.}~\bibnamefont {Mandel}},
  \bibinfo {author} {\bibfnamefont {T.}~\bibnamefont {Esslinger}}, \bibinfo
  {author} {\bibfnamefont {T.~W.}\ \bibnamefont {H\"{a}nsch}}, \ and\ \bibinfo
  {author} {\bibfnamefont {I.}~\bibnamefont {Bloch}},\ }\href {\doibase
  10.1038/415039a} {\bibfield  {journal} {\bibinfo  {journal} {Nature}\
  }\textbf {\bibinfo {volume} {415}},\ \bibinfo {pages} {39} (\bibinfo {year}
  {2002})}\BibitemShut {NoStop}%
\bibitem [{\citenamefont {Fisher}\ \emph {et~al.}(1989)\citenamefont {Fisher},
  \citenamefont {Weichman}, \citenamefont {Grinstein},\ and\ \citenamefont
  {Fisher}}]{fisher}%
  \BibitemOpen
  \bibfield  {author} {\bibinfo {author} {\bibfnamefont {M.~P.~A.}\
  \bibnamefont {Fisher}}, \bibinfo {author} {\bibfnamefont {P.~B.}\
  \bibnamefont {Weichman}}, \bibinfo {author} {\bibfnamefont {G.}~\bibnamefont
  {Grinstein}}, \ and\ \bibinfo {author} {\bibfnamefont {D.~S.}\ \bibnamefont
  {Fisher}},\ }\href {\doibase 10.1103/PhysRevB.40.546} {\bibfield  {journal}
  {\bibinfo  {journal} {Phys. Rev. B}\ }\textbf {\bibinfo {volume} {40}},\
  \bibinfo {pages} {546} (\bibinfo {year} {1989})}\BibitemShut {NoStop}%
\bibitem [{\citenamefont {Greentree}\ \emph {et~al.}(2006)\citenamefont
  {Greentree}, \citenamefont {Tahan}, \citenamefont {Cole},\ and\ \citenamefont
  {Hollenberg}}]{greentree}%
  \BibitemOpen
  \bibfield  {author} {\bibinfo {author} {\bibfnamefont {A.~D.}\ \bibnamefont
  {Greentree}}, \bibinfo {author} {\bibfnamefont {C.}~\bibnamefont {Tahan}},
  \bibinfo {author} {\bibfnamefont {J.~H.}\ \bibnamefont {Cole}}, \ and\
  \bibinfo {author} {\bibfnamefont {L.~C.~L.}\ \bibnamefont {Hollenberg}},\
  }\href {\doibase 10.1038/nphys466} {\bibfield  {journal} {\bibinfo  {journal}
  {Nat. Phys.}\ }\textbf {\bibinfo {volume} {2}},\ \bibinfo {pages} {856}
  (\bibinfo {year} {2006})}\BibitemShut {NoStop}%
\bibitem [{\citenamefont {Hartmann}\ \emph {et~al.}(2006)\citenamefont
  {Hartmann}, \citenamefont {Brand\~ao},\ and\ \citenamefont
  {Plenio}}]{hartmann}%
  \BibitemOpen
  \bibfield  {author} {\bibinfo {author} {\bibfnamefont {M.~J.}\ \bibnamefont
  {Hartmann}}, \bibinfo {author} {\bibfnamefont {F.~G. S.~L.}\ \bibnamefont
  {Brand\~ao}}, \ and\ \bibinfo {author} {\bibfnamefont {M.~B.}\ \bibnamefont
  {Plenio}},\ }\href {\doibase 10.1038/nphys462} {\bibfield  {journal}
  {\bibinfo  {journal} {Nat. Phys.}\ }\textbf {\bibinfo {volume} {2}},\
  \bibinfo {pages} {849} (\bibinfo {year} {2006})}\BibitemShut {NoStop}%
\bibitem [{\citenamefont {Angelakis}\ \emph {et~al.}(2007)\citenamefont
  {Angelakis}, \citenamefont {Santos},\ and\ \citenamefont {Bose}}]{angelakis}%
  \BibitemOpen
  \bibfield  {author} {\bibinfo {author} {\bibfnamefont {D.~G.}\ \bibnamefont
  {Angelakis}}, \bibinfo {author} {\bibfnamefont {M.~F.}\ \bibnamefont
  {Santos}}, \ and\ \bibinfo {author} {\bibfnamefont {S.}~\bibnamefont
  {Bose}},\ }\href {\doibase 10.1103/PhysRevA.76.031805} {\bibfield  {journal}
  {\bibinfo  {journal} {Phys. Rev. A}\ }\textbf {\bibinfo {volume} {76}},\
  \bibinfo {pages} {031805} (\bibinfo {year} {2007})}\BibitemShut {NoStop}%
\bibitem [{\citenamefont {Rossini}\ and\ \citenamefont
  {Fazio}(2007)}]{rossini}%
  \BibitemOpen
  \bibfield  {author} {\bibinfo {author} {\bibfnamefont {D.}~\bibnamefont
  {Rossini}}\ and\ \bibinfo {author} {\bibfnamefont {R.}~\bibnamefont
  {Fazio}},\ }\href {\doibase 10.1103/PhysRevLett.99.186401} {\bibfield
  {journal} {\bibinfo  {journal} {Phys. Rev. Lett.}\ }\textbf {\bibinfo
  {volume} {99}},\ \bibinfo {pages} {186401} (\bibinfo {year}
  {2007})}\BibitemShut {NoStop}%
\bibitem [{\citenamefont {Irish}\ \emph {et~al.}(2008)\citenamefont {Irish},
  \citenamefont {Ogden},\ and\ \citenamefont {Kim}}]{irish}%
  \BibitemOpen
  \bibfield  {author} {\bibinfo {author} {\bibfnamefont {E.~K.}\ \bibnamefont
  {Irish}}, \bibinfo {author} {\bibfnamefont {C.~D.}\ \bibnamefont {Ogden}}, \
  and\ \bibinfo {author} {\bibfnamefont {M.~S.}\ \bibnamefont {Kim}},\ }\href
  {\doibase 10.1103/PhysRevA.77.033801} {\bibfield  {journal} {\bibinfo
  {journal} {Phys. Rev. A}\ }\textbf {\bibinfo {volume} {77}},\ \bibinfo
  {pages} {033801} (\bibinfo {year} {2008})}\BibitemShut {NoStop}%
\bibitem [{\citenamefont {Makin}\ \emph {et~al.}(2008)\citenamefont {Makin},
  \citenamefont {Cole}, \citenamefont {Tahan}, \citenamefont {Hollenberg},\
  and\ \citenamefont {Greentree}}]{makin}%
  \BibitemOpen
  \bibfield  {author} {\bibinfo {author} {\bibfnamefont {M.~I.}\ \bibnamefont
  {Makin}}, \bibinfo {author} {\bibfnamefont {J.~H.}\ \bibnamefont {Cole}},
  \bibinfo {author} {\bibfnamefont {C.}~\bibnamefont {Tahan}}, \bibinfo
  {author} {\bibfnamefont {L.~C.~L.}\ \bibnamefont {Hollenberg}}, \ and\
  \bibinfo {author} {\bibfnamefont {A.~D.}\ \bibnamefont {Greentree}},\ }\href
  {\doibase 10.1103/PhysRevA.77.053819} {\bibfield  {journal} {\bibinfo
  {journal} {Phys. Rev. A}\ }\textbf {\bibinfo {volume} {77}},\ \bibinfo
  {pages} {053819} (\bibinfo {year} {2008})}\BibitemShut {NoStop}%
\bibitem [{\citenamefont {Toyoda}\ \emph {et~al.}(2013)\citenamefont {Toyoda},
  \citenamefont {Matsuno}, \citenamefont {Noguchi}, \citenamefont {Haze},\ and\
  \citenamefont {Urabe}}]{toyoda1}%
  \BibitemOpen
  \bibfield  {author} {\bibinfo {author} {\bibfnamefont {K.}~\bibnamefont
  {Toyoda}}, \bibinfo {author} {\bibfnamefont {Y.}~\bibnamefont {Matsuno}},
  \bibinfo {author} {\bibfnamefont {A.}~\bibnamefont {Noguchi}}, \bibinfo
  {author} {\bibfnamefont {S.}~\bibnamefont {Haze}}, \ and\ \bibinfo {author}
  {\bibfnamefont {S.}~\bibnamefont {Urabe}},\ }\href {\doibase
  10.1103/PhysRevLett.111.160501} {\bibfield  {journal} {\bibinfo  {journal}
  {Phys. Rev. Lett.}\ }\textbf {\bibinfo {volume} {111}},\ \bibinfo {pages}
  {160501} (\bibinfo {year} {2013})}\BibitemShut {NoStop}%
\bibitem [{\citenamefont {Cai}\ \emph {et~al.}(2021)\citenamefont {Cai},
  \citenamefont {Liu}, \citenamefont {Zhao}, \citenamefont {Wu}, \citenamefont
  {Mei}, \citenamefont {Jiang}, \citenamefont {He}, \citenamefont {Zhang},
  \citenamefont {Zhou},\ and\ \citenamefont {Duan}}]{mlcai_2021}%
  \BibitemOpen
  \bibfield  {author} {\bibinfo {author} {\bibfnamefont {M.~L.}\ \bibnamefont
  {Cai}}, \bibinfo {author} {\bibfnamefont {Z.~D.}\ \bibnamefont {Liu}},
  \bibinfo {author} {\bibfnamefont {W.~D.}\ \bibnamefont {Zhao}}, \bibinfo
  {author} {\bibfnamefont {Y.~K.}\ \bibnamefont {Wu}}, \bibinfo {author}
  {\bibfnamefont {Q.~X.}\ \bibnamefont {Mei}}, \bibinfo {author} {\bibfnamefont
  {Y.}~\bibnamefont {Jiang}}, \bibinfo {author} {\bibfnamefont
  {L.}~\bibnamefont {He}}, \bibinfo {author} {\bibfnamefont {X.}~\bibnamefont
  {Zhang}}, \bibinfo {author} {\bibfnamefont {Z.~C.}\ \bibnamefont {Zhou}}, \
  and\ \bibinfo {author} {\bibfnamefont {L.~M.}\ \bibnamefont {Duan}},\ }\href
  {\doibase 10.1038/s41467-021-21425-8} {\bibfield  {journal} {\bibinfo
  {journal} {Nat. Commun}\ }\textbf {\bibinfo {volume} {12}},\ \bibinfo {pages}
  {1126} (\bibinfo {year} {2021})}\BibitemShut {NoStop}%
\bibitem [{\citenamefont {Cirac}\ and\ \citenamefont {Zoller}(1995)}]{cirac}%
  \BibitemOpen
  \bibfield  {author} {\bibinfo {author} {\bibfnamefont {J.~I.}\ \bibnamefont
  {Cirac}}\ and\ \bibinfo {author} {\bibfnamefont {P.}~\bibnamefont {Zoller}},\
  }\href {\doibase 10.1103/PhysRevLett.74.4091} {\bibfield  {journal} {\bibinfo
   {journal} {Phys. Rev. Lett.}\ }\textbf {\bibinfo {volume} {74}},\ \bibinfo
  {pages} {4091} (\bibinfo {year} {1995})}\BibitemShut {NoStop}%
\bibitem [{\citenamefont {Toyoda}\ \emph {et~al.}(2010)\citenamefont {Toyoda},
  \citenamefont {Haze}, \citenamefont {Yamazaki},\ and\ \citenamefont
  {Urabe}}]{toyoda2}%
  \BibitemOpen
  \bibfield  {author} {\bibinfo {author} {\bibfnamefont {K.}~\bibnamefont
  {Toyoda}}, \bibinfo {author} {\bibfnamefont {S.}~\bibnamefont {Haze}},
  \bibinfo {author} {\bibfnamefont {R.}~\bibnamefont {Yamazaki}}, \ and\
  \bibinfo {author} {\bibfnamefont {S.}~\bibnamefont {Urabe}},\ }\href
  {\doibase 10.1103/PhysRevA.81.032322} {\bibfield  {journal} {\bibinfo
  {journal} {Phys. Rev. A}\ }\textbf {\bibinfo {volume} {81}},\ \bibinfo
  {pages} {032322} (\bibinfo {year} {2010})}\BibitemShut {NoStop}%
\bibitem [{\citenamefont {Leibfried}\ \emph {et~al.}(2003)\citenamefont
  {Leibfried}, \citenamefont {Blatt}, \citenamefont {Monroe},\ and\
  \citenamefont {Wineland}}]{leibfried03}%
  \BibitemOpen
  \bibfield  {author} {\bibinfo {author} {\bibfnamefont {D.}~\bibnamefont
  {Leibfried}}, \bibinfo {author} {\bibfnamefont {R.}~\bibnamefont {Blatt}},
  \bibinfo {author} {\bibfnamefont {C.}~\bibnamefont {Monroe}}, \ and\ \bibinfo
  {author} {\bibfnamefont {D.}~\bibnamefont {Wineland}},\ }\href {\doibase
  10.1103/RevModPhys.75.281} {\bibfield  {journal} {\bibinfo  {journal} {Rev.
  Mod. Phys.}\ }\textbf {\bibinfo {volume} {75}},\ \bibinfo {pages} {281}
  (\bibinfo {year} {2003})}\BibitemShut {NoStop}%
\bibitem [{\citenamefont {Hays}\ \emph {et~al.}(2020)\citenamefont {Hays},
  \citenamefont {Fatemi}, \citenamefont {Serniak}, \citenamefont {Bouman},
  \citenamefont {Diamond}, \citenamefont {de~Lange}, \citenamefont {Krogstrup},
  \citenamefont {Nygård}, \citenamefont {Geresdi},\ and\ \citenamefont
  {Devoret}}]{hays_continuous_2020}%
  \BibitemOpen
  \bibfield  {author} {\bibinfo {author} {\bibfnamefont {M.}~\bibnamefont
  {Hays}}, \bibinfo {author} {\bibfnamefont {V.}~\bibnamefont {Fatemi}},
  \bibinfo {author} {\bibfnamefont {K.}~\bibnamefont {Serniak}}, \bibinfo
  {author} {\bibfnamefont {D.}~\bibnamefont {Bouman}}, \bibinfo {author}
  {\bibfnamefont {S.}~\bibnamefont {Diamond}}, \bibinfo {author} {\bibfnamefont
  {G.}~\bibnamefont {de~Lange}}, \bibinfo {author} {\bibfnamefont
  {P.}~\bibnamefont {Krogstrup}}, \bibinfo {author} {\bibfnamefont
  {J.}~\bibnamefont {Nygård}}, \bibinfo {author} {\bibfnamefont
  {A.}~\bibnamefont {Geresdi}}, \ and\ \bibinfo {author} {\bibfnamefont
  {M.~H.}\ \bibnamefont {Devoret}},\ }\href {\doibase
  10.1038/s41567-020-0952-3} {\bibfield  {journal} {\bibinfo  {journal} {Nat.
  Phys.}\ }\textbf {\bibinfo {volume} {16}},\ \bibinfo {pages} {1103} (\bibinfo
  {year} {2020})}\BibitemShut {NoStop}%
\bibitem [{\citenamefont {Porras}\ and\ \citenamefont {Cirac}(2004)}]{porras}%
  \BibitemOpen
  \bibfield  {author} {\bibinfo {author} {\bibfnamefont {D.}~\bibnamefont
  {Porras}}\ and\ \bibinfo {author} {\bibfnamefont {J.~I.}\ \bibnamefont
  {Cirac}},\ }\href {\doibase 10.1103/PhysRevLett.93.263602} {\bibfield
  {journal} {\bibinfo  {journal} {Phys. Rev. Lett.}\ }\textbf {\bibinfo
  {volume} {93}},\ \bibinfo {pages} {263602} (\bibinfo {year}
  {2004})}\BibitemShut {NoStop}%
\bibitem [{\citenamefont {Deng}\ \emph {et~al.}(2008)\citenamefont {Deng},
  \citenamefont {Porras},\ and\ \citenamefont {Cirac}}]{deng}%
  \BibitemOpen
  \bibfield  {author} {\bibinfo {author} {\bibfnamefont {X.-L.}\ \bibnamefont
  {Deng}}, \bibinfo {author} {\bibfnamefont {D.}~\bibnamefont {Porras}}, \ and\
  \bibinfo {author} {\bibfnamefont {J.~I.}\ \bibnamefont {Cirac}},\ }\href
  {\doibase 10.1103/PhysRevA.77.033403} {\bibfield  {journal} {\bibinfo
  {journal} {Phys. Rev. A}\ }\textbf {\bibinfo {volume} {77}},\ \bibinfo
  {pages} {033403} (\bibinfo {year} {2008})}\BibitemShut {NoStop}%
\bibitem [{\citenamefont {Wineland}\ \emph {et~al.}(1998)\citenamefont
  {Wineland}, \citenamefont {Monroe}, \citenamefont {Itano}, \citenamefont
  {Leibfried}, \citenamefont {King},\ and\ \citenamefont {Meekhof}}]{wineland}%
  \BibitemOpen
  \bibfield  {author} {\bibinfo {author} {\bibfnamefont {D.~J.}\ \bibnamefont
  {Wineland}}, \bibinfo {author} {\bibfnamefont {C.}~\bibnamefont {Monroe}},
  \bibinfo {author} {\bibfnamefont {W.~M.}\ \bibnamefont {Itano}}, \bibinfo
  {author} {\bibfnamefont {D.}~\bibnamefont {Leibfried}}, \bibinfo {author}
  {\bibfnamefont {B.~E.}\ \bibnamefont {King}}, \ and\ \bibinfo {author}
  {\bibfnamefont {D.~M.}\ \bibnamefont {Meekhof}},\ }\href {\doibase
  10.6028/jres.103.019} {\bibfield  {journal} {\bibinfo  {journal} {J. Res.
  Natl. Inst. Stand. Technol.}\ }\textbf {\bibinfo {volume} {103}},\ \bibinfo
  {pages} {259} (\bibinfo {year} {1998})}\BibitemShut {NoStop}%
\bibitem [{\citenamefont {Vidal}\ and\ \citenamefont {Werner}(2002)}]{vidal}%
  \BibitemOpen
  \bibfield  {author} {\bibinfo {author} {\bibfnamefont {G.}~\bibnamefont
  {Vidal}}\ and\ \bibinfo {author} {\bibfnamefont {R.~F.}\ \bibnamefont
  {Werner}},\ }\href {\doibase 10.1103/PhysRevA.65.032314} {\bibfield
  {journal} {\bibinfo  {journal} {Phys. Rev. A}\ }\textbf {\bibinfo {volume}
  {65}},\ \bibinfo {pages} {032314} (\bibinfo {year} {2002})}\BibitemShut
  {NoStop}%
\bibitem [{\citenamefont {Plenio}(2005)}]{plenio}%
  \BibitemOpen
  \bibfield  {author} {\bibinfo {author} {\bibfnamefont {M.~B.}\ \bibnamefont
  {Plenio}},\ }\href {\doibase 10.1103/PhysRevLett.95.090503} {\bibfield
  {journal} {\bibinfo  {journal} {Phys. Rev. Lett.}\ }\textbf {\bibinfo
  {volume} {95}},\ \bibinfo {pages} {090503} (\bibinfo {year}
  {2005})}\BibitemShut {NoStop}%
\bibitem [{\citenamefont {Haze}\ \emph {et~al.}(2012)\citenamefont {Haze},
  \citenamefont {Tateishi}, \citenamefont {Noguchi}, \citenamefont {Toyoda},\
  and\ \citenamefont {Urabe}}]{haze2012}%
  \BibitemOpen
  \bibfield  {author} {\bibinfo {author} {\bibfnamefont {S.}~\bibnamefont
  {Haze}}, \bibinfo {author} {\bibfnamefont {Y.}~\bibnamefont {Tateishi}},
  \bibinfo {author} {\bibfnamefont {A.}~\bibnamefont {Noguchi}}, \bibinfo
  {author} {\bibfnamefont {K.}~\bibnamefont {Toyoda}}, \ and\ \bibinfo {author}
  {\bibfnamefont {S.}~\bibnamefont {Urabe}},\ }\href {\doibase
  10.1103/PhysRevA.85.031401} {\bibfield  {journal} {\bibinfo  {journal} {Phys.
  Rev. A}\ }\textbf {\bibinfo {volume} {85}},\ \bibinfo {pages} {031401}
  (\bibinfo {year} {2012})}\BibitemShut {NoStop}%
\bibitem [{\citenamefont {Toyoda}\ \emph {et~al.}(2015)\citenamefont {Toyoda},
  \citenamefont {Hiji}, \citenamefont {Noguchi},\ and\ \citenamefont
  {Urabe}}]{toyoda2015}%
  \BibitemOpen
  \bibfield  {author} {\bibinfo {author} {\bibfnamefont {K.}~\bibnamefont
  {Toyoda}}, \bibinfo {author} {\bibfnamefont {R.}~\bibnamefont {Hiji}},
  \bibinfo {author} {\bibfnamefont {A.}~\bibnamefont {Noguchi}}, \ and\
  \bibinfo {author} {\bibfnamefont {S.}~\bibnamefont {Urabe}},\ }\href
  {\doibase 10.1038/nature15735} {\bibfield  {journal} {\bibinfo  {journal}
  {Nature}\ }\textbf {\bibinfo {volume} {527}},\ \bibinfo {pages} {74}
  (\bibinfo {year} {2015})}\BibitemShut {NoStop}%
\bibitem [{\citenamefont {Zhang}\ \emph {et~al.}(2018)\citenamefont {Zhang},
  \citenamefont {Um}, \citenamefont {Lv}, \citenamefont {Zhang}, \citenamefont
  {Duan},\ and\ \citenamefont {Kim}}]{zhang2018}%
  \BibitemOpen
  \bibfield  {author} {\bibinfo {author} {\bibfnamefont {J.}~\bibnamefont
  {Zhang}}, \bibinfo {author} {\bibfnamefont {M.}~\bibnamefont {Um}}, \bibinfo
  {author} {\bibfnamefont {D.}~\bibnamefont {Lv}}, \bibinfo {author}
  {\bibfnamefont {J.-N.}\ \bibnamefont {Zhang}}, \bibinfo {author}
  {\bibfnamefont {L.-M.}\ \bibnamefont {Duan}}, \ and\ \bibinfo {author}
  {\bibfnamefont {K.}~\bibnamefont {Kim}},\ }\href {\doibase
  10.1103/PhysRevLett.121.160502} {\bibfield  {journal} {\bibinfo  {journal}
  {Phys. Rev. Lett.}\ }\textbf {\bibinfo {volume} {121}},\ \bibinfo {pages}
  {160502} (\bibinfo {year} {2018})}\BibitemShut {NoStop}%
\bibitem [{\citenamefont {Tavis}\ and\ \citenamefont
  {Cummings}(1968)}]{tavis_exact_1968}%
  \BibitemOpen
  \bibfield  {author} {\bibinfo {author} {\bibfnamefont {M.}~\bibnamefont
  {Tavis}}\ and\ \bibinfo {author} {\bibfnamefont {F.~W.}\ \bibnamefont
  {Cummings}},\ }\href {\doibase 10.1103/PhysRev.170.379} {\bibfield  {journal}
  {\bibinfo  {journal} {Phys. Rev.}\ }\textbf {\bibinfo {volume} {170}},\
  \bibinfo {pages} {379} (\bibinfo {year} {1968})}\BibitemShut {NoStop}%
\bibitem [{\citenamefont {Tavis}\ and\ \citenamefont
  {Cummings}(1969)}]{tavis_approximate_1969}%
  \BibitemOpen
  \bibfield  {author} {\bibinfo {author} {\bibfnamefont {M.}~\bibnamefont
  {Tavis}}\ and\ \bibinfo {author} {\bibfnamefont {F.~W.}\ \bibnamefont
  {Cummings}},\ }\href {\doibase 10.1103/PhysRev.188.692} {\bibfield  {journal}
  {\bibinfo  {journal} {Phys. Rev.}\ }\textbf {\bibinfo {volume} {188}},\
  \bibinfo {pages} {692} (\bibinfo {year} {1969})}\BibitemShut {NoStop}%
\bibitem [{\citenamefont {Kirton}\ \emph {et~al.}(2019)\citenamefont {Kirton},
  \citenamefont {Roses}, \citenamefont {Keeling},\ and\ \citenamefont
  {Torre}}]{kirton_introduction_2019}%
  \BibitemOpen
  \bibfield  {author} {\bibinfo {author} {\bibfnamefont {P.}~\bibnamefont
  {Kirton}}, \bibinfo {author} {\bibfnamefont {M.~M.}\ \bibnamefont {Roses}},
  \bibinfo {author} {\bibfnamefont {J.}~\bibnamefont {Keeling}}, \ and\
  \bibinfo {author} {\bibfnamefont {E.~G.~D.}\ \bibnamefont {Torre}},\ }\href
  {\doibase https://doi.org/10.1002/qute.201800043} {\bibfield  {journal}
  {\bibinfo  {journal} {Adv. Quantum Technol.}\ }\textbf {\bibinfo {volume}
  {2}},\ \bibinfo {pages} {1800043} (\bibinfo {year} {2019})}\BibitemShut
  {NoStop}%
\bibitem [{\citenamefont {Aedo}\ and\ \citenamefont
  {Lamata}(2018)}]{aedo_analog_2018}%
  \BibitemOpen
  \bibfield  {author} {\bibinfo {author} {\bibfnamefont {I.}~\bibnamefont
  {Aedo}}\ and\ \bibinfo {author} {\bibfnamefont {L.}~\bibnamefont {Lamata}},\
  }\href {\doibase 10.1103/PhysRevA.97.042317} {\bibfield  {journal} {\bibinfo
  {journal} {Phys. Rev. A}\ }\textbf {\bibinfo {volume} {97}},\ \bibinfo
  {pages} {042317} (\bibinfo {year} {2018})}\BibitemShut {NoStop}%
\bibitem [{\citenamefont {Kirton}\ and\ \citenamefont
  {Keeling}(2018)}]{kirton_superradiant_2018}%
  \BibitemOpen
  \bibfield  {author} {\bibinfo {author} {\bibfnamefont {P.}~\bibnamefont
  {Kirton}}\ and\ \bibinfo {author} {\bibfnamefont {J.}~\bibnamefont
  {Keeling}},\ }\href {\doibase 10.1088/1367-2630/aaa11d} {\bibfield  {journal}
  {\bibinfo  {journal} {New J. Phys.}\ }\textbf {\bibinfo {volume} {20}},\
  \bibinfo {pages} {015009} (\bibinfo {year} {2018})}\BibitemShut {NoStop}%
\bibitem [{\citenamefont {Kim}\ \emph {et~al.}(2007)\citenamefont {Kim},
  \citenamefont {Palma},\ and\ \citenamefont
  {Paternostro}}]{kim_accumulation_2007}%
  \BibitemOpen
  \bibfield  {author} {\bibinfo {author} {\bibfnamefont {M.~S.}\ \bibnamefont
  {Kim}}, \bibinfo {author} {\bibfnamefont {G.~M.}\ \bibnamefont {Palma}}, \
  and\ \bibinfo {author} {\bibfnamefont {M.}~\bibnamefont {Paternostro}},\
  }\href {\doibase 10.1103/PhysRevLett.98.140504} {\bibfield  {journal}
  {\bibinfo  {journal} {Phys. Rev. Lett.}\ }\textbf {\bibinfo {volume} {98}},\
  \bibinfo {pages} {140504} (\bibinfo {year} {2007})}\BibitemShut {NoStop}%
\bibitem [{\citenamefont {Cirac}\ \emph {et~al.}(1994)\citenamefont {Cirac},
  \citenamefont {Blatt}, \citenamefont {Parkins},\ and\ \citenamefont
  {Zoller}}]{cirac_pra_1994}%
  \BibitemOpen
  \bibfield  {author} {\bibinfo {author} {\bibfnamefont {J.~I.}\ \bibnamefont
  {Cirac}}, \bibinfo {author} {\bibfnamefont {R.}~\bibnamefont {Blatt}},
  \bibinfo {author} {\bibfnamefont {A.~S.}\ \bibnamefont {Parkins}}, \ and\
  \bibinfo {author} {\bibfnamefont {P.}~\bibnamefont {Zoller}},\ }\href
  {\doibase 10.1103/PhysRevA.49.1202} {\bibfield  {journal} {\bibinfo
  {journal} {Phys. Rev. A}\ }\textbf {\bibinfo {volume} {49}},\ \bibinfo
  {pages} {1202} (\bibinfo {year} {1994})}\BibitemShut {NoStop}%
\bibitem [{\citenamefont {Myerson}\ \emph {et~al.}(2008)\citenamefont
  {Myerson}, \citenamefont {Szwer}, \citenamefont {Webster}, \citenamefont
  {Allcock}, \citenamefont {Curtis}, \citenamefont {Imreh}, \citenamefont
  {Sherman}, \citenamefont {Stacey}, \citenamefont {Steane},\ and\
  \citenamefont {Lucas}}]{myerson_high-fidelity_2008}%
  \BibitemOpen
  \bibfield  {author} {\bibinfo {author} {\bibfnamefont {A.~H.}\ \bibnamefont
  {Myerson}}, \bibinfo {author} {\bibfnamefont {D.~J.}\ \bibnamefont {Szwer}},
  \bibinfo {author} {\bibfnamefont {S.~C.}\ \bibnamefont {Webster}}, \bibinfo
  {author} {\bibfnamefont {D.~T.~C.}\ \bibnamefont {Allcock}}, \bibinfo
  {author} {\bibfnamefont {M.~J.}\ \bibnamefont {Curtis}}, \bibinfo {author}
  {\bibfnamefont {G.}~\bibnamefont {Imreh}}, \bibinfo {author} {\bibfnamefont
  {J.~A.}\ \bibnamefont {Sherman}}, \bibinfo {author} {\bibfnamefont {D.~N.}\
  \bibnamefont {Stacey}}, \bibinfo {author} {\bibfnamefont {A.~M.}\
  \bibnamefont {Steane}}, \ and\ \bibinfo {author} {\bibfnamefont {D.~M.}\
  \bibnamefont {Lucas}},\ }\href {\doibase 10.1103/PhysRevLett.100.200502}
  {\bibfield  {journal} {\bibinfo  {journal} {Phys. Rev. Lett.}\ }\textbf
  {\bibinfo {volume} {100}},\ \bibinfo {pages} {200502} (\bibinfo {year}
  {2008})}\BibitemShut {NoStop}%
\bibitem [{\citenamefont {Paternostro}\ \emph {et~al.}(2006)\citenamefont
  {Paternostro}, \citenamefont {Kim}, \citenamefont {Bose},\ and\ \citenamefont
  {Lee}}]{paternostro_entanglement_2006}%
  \BibitemOpen
  \bibfield  {author} {\bibinfo {author} {\bibfnamefont {M.}~\bibnamefont
  {Paternostro}}, \bibinfo {author} {\bibfnamefont {M.~S.}\ \bibnamefont
  {Kim}}, \bibinfo {author} {\bibfnamefont {S.}~\bibnamefont {Bose}}, \ and\
  \bibinfo {author} {\bibfnamefont {J.}~\bibnamefont {Lee}},\ }\href {\doibase
  10.1103/PhysRevLett.96.080501} {\bibfield  {journal} {\bibinfo  {journal}
  {Phys. Rev. Lett.}\ }\textbf {\bibinfo {volume} {96}},\ \bibinfo {pages}
  {080501} (\bibinfo {year} {2006})}\BibitemShut {NoStop}%
\bibitem [{\citenamefont {Adesso}\ and\ \citenamefont
  {Illuminati}(2008)}]{adesso_genuine_2008}%
  \BibitemOpen
  \bibfield  {author} {\bibinfo {author} {\bibfnamefont {G.}~\bibnamefont
  {Adesso}}\ and\ \bibinfo {author} {\bibfnamefont {F.}~\bibnamefont
  {Illuminati}},\ }\href {\doibase 10.1103/PhysRevA.78.042310} {\bibfield
  {journal} {\bibinfo  {journal} {Phys. Rev. A}\ }\textbf {\bibinfo {volume}
  {78}},\ \bibinfo {pages} {042310} (\bibinfo {year} {2008})}\BibitemShut
  {NoStop}%
\bibitem [{\citenamefont {Podhora}\ \emph {et~al.}(2020)\citenamefont
  {Podhora}, \citenamefont {Pham}, \citenamefont {Lešundák}, \citenamefont
  {Obšil}, \citenamefont {Čížek}, \citenamefont {Číp}, \citenamefont
  {Marek}, \citenamefont {Slodička},\ and\ \citenamefont
  {Filip}}]{podhora_unconditional_2020}%
  \BibitemOpen
  \bibfield  {author} {\bibinfo {author} {\bibfnamefont {L.}~\bibnamefont
  {Podhora}}, \bibinfo {author} {\bibfnamefont {T.}~\bibnamefont {Pham}},
  \bibinfo {author} {\bibfnamefont {A.}~\bibnamefont {Lešundák}}, \bibinfo
  {author} {\bibfnamefont {P.}~\bibnamefont {Obšil}}, \bibinfo {author}
  {\bibfnamefont {M.}~\bibnamefont {Čížek}}, \bibinfo {author}
  {\bibfnamefont {O.}~\bibnamefont {Číp}}, \bibinfo {author} {\bibfnamefont
  {P.}~\bibnamefont {Marek}}, \bibinfo {author} {\bibfnamefont
  {L.}~\bibnamefont {Slodička}}, \ and\ \bibinfo {author} {\bibfnamefont
  {R.}~\bibnamefont {Filip}},\ }\href {\doibase 10.1002/qute.202000012}
  {\bibfield  {journal} {\bibinfo  {journal} {Adv. Quantum Technol.}\ }\textbf
  {\bibinfo {volume} {3}},\ \bibinfo {pages} {2000012} (\bibinfo {year}
  {2020})}\BibitemShut {NoStop}%
\bibitem [{\citenamefont {Peres}(2002)}]{peres}%
  \BibitemOpen
  \bibfield  {author} {\bibinfo {author} {\bibfnamefont {A.}~\bibnamefont
  {Peres}},\ }\href@noop {} {\emph {\bibinfo {title} {Quantum Theory: Concepts
  and Methods}}}\ (\bibinfo  {publisher} {Springer, Netherlands},\ \bibinfo
  {year} {2002})\BibitemShut {NoStop}%
\bibitem [{\citenamefont {Golter}\ \emph {et~al.}(2016)\citenamefont {Golter},
  \citenamefont {Oo}, \citenamefont {Amezcua}, \citenamefont {Stewart},\ and\
  \citenamefont {Wang}}]{golter}%
  \BibitemOpen
  \bibfield  {author} {\bibinfo {author} {\bibfnamefont {D.~A.}\ \bibnamefont
  {Golter}}, \bibinfo {author} {\bibfnamefont {T.}~\bibnamefont {Oo}}, \bibinfo
  {author} {\bibfnamefont {M.}~\bibnamefont {Amezcua}}, \bibinfo {author}
  {\bibfnamefont {K.~A.}\ \bibnamefont {Stewart}}, \ and\ \bibinfo {author}
  {\bibfnamefont {H.}~\bibnamefont {Wang}},\ }\href {\doibase
  10.1103/PhysRevLett.116.143602} {\bibfield  {journal} {\bibinfo  {journal}
  {Phys. Rev. Lett.}\ }\textbf {\bibinfo {volume} {116}},\ \bibinfo {pages}
  {143602} (\bibinfo {year} {2016})}\BibitemShut {NoStop}%
\bibitem [{\citenamefont {Gärtner}\ \emph {et~al.}(2018)\citenamefont
  {Gärtner}, \citenamefont {Moura}, \citenamefont {Haaxman}, \citenamefont
  {Norte},\ and\ \citenamefont {Gröblacher}}]{gartner}%
  \BibitemOpen
  \bibfield  {author} {\bibinfo {author} {\bibfnamefont {C.}~\bibnamefont
  {Gärtner}}, \bibinfo {author} {\bibfnamefont {J.~P.}\ \bibnamefont {Moura}},
  \bibinfo {author} {\bibfnamefont {W.}~\bibnamefont {Haaxman}}, \bibinfo
  {author} {\bibfnamefont {R.~A.}\ \bibnamefont {Norte}}, \ and\ \bibinfo
  {author} {\bibfnamefont {S.}~\bibnamefont {Gröblacher}},\ }\href {\doibase
  10.1021/acs.nanolett.8b03240} {\bibfield  {journal} {\bibinfo  {journal}
  {Nano Lett.}\ }\textbf {\bibinfo {volume} {18}},\ \bibinfo {pages} {7171}
  (\bibinfo {year} {2018})}\BibitemShut {NoStop}%
\bibitem [{\citenamefont {Arita}\ \emph {et~al.}(2018)\citenamefont {Arita},
  \citenamefont {Wright},\ and\ \citenamefont {Dholakia}}]{arita}%
  \BibitemOpen
  \bibfield  {author} {\bibinfo {author} {\bibfnamefont {Y.}~\bibnamefont
  {Arita}}, \bibinfo {author} {\bibfnamefont {E.~M.}\ \bibnamefont {Wright}}, \
  and\ \bibinfo {author} {\bibfnamefont {K.}~\bibnamefont {Dholakia}},\ }\href
  {\doibase 10.1364/OPTICA.5.000910} {\bibfield  {journal} {\bibinfo  {journal}
  {Optica}\ }\textbf {\bibinfo {volume} {5}},\ \bibinfo {pages} {910} (\bibinfo
  {year} {2018})}\BibitemShut {NoStop}%
\bibitem [{\citenamefont {Yang}\ \emph {et~al.}(2020)\citenamefont {Yang},
  \citenamefont {Wei}, \citenamefont {Sheng},\ and\ \citenamefont
  {Wu}}]{cyang}%
  \BibitemOpen
  \bibfield  {author} {\bibinfo {author} {\bibfnamefont {C.}~\bibnamefont
  {Yang}}, \bibinfo {author} {\bibfnamefont {X.}~\bibnamefont {Wei}}, \bibinfo
  {author} {\bibfnamefont {J.}~\bibnamefont {Sheng}}, \ and\ \bibinfo {author}
  {\bibfnamefont {H.}~\bibnamefont {Wu}},\ }\href {\doibase
  10.1038/s41467-020-18426-4} {\bibfield  {journal} {\bibinfo  {journal} {Nat.
  Commun.}\ }\textbf {\bibinfo {volume} {11}},\ \bibinfo {pages} {4656}
  (\bibinfo {year} {2020})}\BibitemShut {NoStop}%
\bibitem [{\citenamefont {Rogers}\ \emph {et~al.}(2014)\citenamefont {Rogers},
  \citenamefont {Gullo}, \citenamefont {Chiara}, \citenamefont {Palma},\ and\
  \citenamefont {Paternostro}}]{rogers_hybrid_2014}%
  \BibitemOpen
  \bibfield  {author} {\bibinfo {author} {\bibfnamefont {B.}~\bibnamefont
  {Rogers}}, \bibinfo {author} {\bibfnamefont {N.~L.}\ \bibnamefont {Gullo}},
  \bibinfo {author} {\bibfnamefont {G.~D.}\ \bibnamefont {Chiara}}, \bibinfo
  {author} {\bibfnamefont {G.~M.}\ \bibnamefont {Palma}}, \ and\ \bibinfo
  {author} {\bibfnamefont {M.}~\bibnamefont {Paternostro}},\ }\href {\doibase
  doi:10.2478/qmetro-2014-0002} {\bibfield  {journal} {\bibinfo  {journal}
  {Quantum Meas. Quantum Metrol.}\ }\textbf {\bibinfo {volume} {2}},\ \bibinfo
  {pages} {11} (\bibinfo {year} {2014})}\BibitemShut {NoStop}%
\bibitem [{\citenamefont {Andersen}\ \emph {et~al.}(2015)\citenamefont
  {Andersen}, \citenamefont {Neergaard-Nielsen}, \citenamefont {van Loock},\
  and\ \citenamefont {Furusawa}}]{andersen_hybrid_2015}%
  \BibitemOpen
  \bibfield  {author} {\bibinfo {author} {\bibfnamefont {U.~L.}\ \bibnamefont
  {Andersen}}, \bibinfo {author} {\bibfnamefont {J.~S.}\ \bibnamefont
  {Neergaard-Nielsen}}, \bibinfo {author} {\bibfnamefont {P.}~\bibnamefont {van
  Loock}}, \ and\ \bibinfo {author} {\bibfnamefont {A.}~\bibnamefont
  {Furusawa}},\ }\href {\doibase 10.1038/nphys3410} {\bibfield  {journal}
  {\bibinfo  {journal} {Nat. Phys.}\ }\textbf {\bibinfo {volume} {11}},\
  \bibinfo {pages} {713} (\bibinfo {year} {2015})}\BibitemShut {NoStop}%
\bibitem [{\citenamefont {Chu}\ \emph {et~al.}(2017)\citenamefont {Chu},
  \citenamefont {Kharel}, \citenamefont {Renninger}, \citenamefont {Burkhart},
  \citenamefont {Frunzio}, \citenamefont {Rakich},\ and\ \citenamefont
  {Schoelkopf}}]{chu_quantum_2017}%
  \BibitemOpen
  \bibfield  {author} {\bibinfo {author} {\bibfnamefont {Y.}~\bibnamefont
  {Chu}}, \bibinfo {author} {\bibfnamefont {P.}~\bibnamefont {Kharel}},
  \bibinfo {author} {\bibfnamefont {W.~H.}\ \bibnamefont {Renninger}}, \bibinfo
  {author} {\bibfnamefont {L.~D.}\ \bibnamefont {Burkhart}}, \bibinfo {author}
  {\bibfnamefont {L.}~\bibnamefont {Frunzio}}, \bibinfo {author} {\bibfnamefont
  {P.~T.}\ \bibnamefont {Rakich}}, \ and\ \bibinfo {author} {\bibfnamefont
  {R.~J.}\ \bibnamefont {Schoelkopf}},\ }\href {\doibase
  10.1126/science.aao1511} {\bibfield  {journal} {\bibinfo  {journal}
  {Science}\ }\textbf {\bibinfo {volume} {358}},\ \bibinfo {pages} {199}
  (\bibinfo {year} {2017})}\BibitemShut {NoStop}%
\bibitem [{\citenamefont {Cady}\ \emph {et~al.}(2019)\citenamefont {Cady},
  \citenamefont {Michel}, \citenamefont {Lee}, \citenamefont {Patel},
  \citenamefont {Sarabalis}, \citenamefont {Safavi-Naeini},\ and\ \citenamefont
  {Jayich}}]{cady_diamond_2019}%
  \BibitemOpen
  \bibfield  {author} {\bibinfo {author} {\bibfnamefont {J.~V.}\ \bibnamefont
  {Cady}}, \bibinfo {author} {\bibfnamefont {O.}~\bibnamefont {Michel}},
  \bibinfo {author} {\bibfnamefont {K.~W.}\ \bibnamefont {Lee}}, \bibinfo
  {author} {\bibfnamefont {R.~N.}\ \bibnamefont {Patel}}, \bibinfo {author}
  {\bibfnamefont {C.~J.}\ \bibnamefont {Sarabalis}}, \bibinfo {author}
  {\bibfnamefont {A.~H.}\ \bibnamefont {Safavi-Naeini}}, \ and\ \bibinfo
  {author} {\bibfnamefont {A.~C.~B.}\ \bibnamefont {Jayich}},\ }\href {\doibase
  10.1088/2058-9565/ab043e} {\bibfield  {journal} {\bibinfo  {journal} {Quantum
  Sci. Technol.}\ }\textbf {\bibinfo {volume} {4}},\ \bibinfo {pages} {024009}
  (\bibinfo {year} {2019})}\BibitemShut {NoStop}%
\end{thebibliography}%


\widetext
\newpage
\begin{center}
\textbf{\large Supplementary Material}
\end{center}
\setcounter{equation}{0}
\setcounter{figure}{0}
\setcounter{table}{0}
\setcounter{page}{1}
\setcounter{section}{0}
\makeatletter
\renewcommand{\theequation}{S\arabic{equation}}
\renewcommand{\thefigure}{S\arabic{figure}}
\renewcommand{\bibnumfmt}[1]{[S#1]}
\renewcommand{\citenumfont}[1]{S#1}

\section{Derivation of the radial Hamiltonian}
In a typical rf or Paul trap, the trapping potential is taken to be harmonic in all three spatial directions ($\alpha=x,y,z$) with respective secular frequencies $\omega_x$, $\omega_y$ and $\omega_z$. Therefore, for a single ion with mass $m$ and charge $e$, the effective trapping potential  is  $V_{\text{tr}} = \sum_{\alpha} \tfrac{1}{2}m \omega_{\alpha}^{2} \alpha^{2}$.

For a pair of trapped ions with equal mass and charge, the effective potential energy $V$ has an extra Coulombic contribution $V_C$, i.e., $V = V_{\text{tr}} + V_C$, where 
\begin{equation}
   V_C = \frac{e^{2}}{\vert \mathbf{r}_1 - \mathbf{r}_2\vert} =    \frac{e^2}{\sqrt{(x_1 - x_2)^2+(y_1 - y_2)^2 + (z_1 - z_2)^2}} \,,
\end{equation}
where $\mathbf{r}_i = (x_i, y_i, z_i)$ with $i=1,\,2$. Since each ion is identical and the trap is assumed to be homogeneous, the secular frequencies are also identical for each ion. Since there are two ions, we can simply choose the $z$-axis to be along the line connecting the ions and solve for the $z$-axis equilibrium points by setting $\frac{\partial V}{\partial z_i}=0$. In general, if the trapping potential is arranged such that $\omega_x,\omega_y\gg \omega_z$, then the ions will distribute along the $z$-axis, with equilibrium positions $(0, 0, z_{i}^{0})$, where $z_1^0 = -\sqrt[3]{\frac{e^2}{4m\omega_z^2}}=-z_2^0$. $V_C$ can be rewritten in terms of the displacements from equilibrium and the motion momentarily collected into relative motions $X=x_1-x_2$, $Y=y_1-y_2$ and $Z=z_1-z_2$. The equilibrium position of $Z$ is inherited from the local coordinates as $Z^0=z_1^0-z_2^0$. Then $V_C$ is expressed as
\begin{equation}
    V_C=\frac{e^{2}}{\sqrt{X^2+Y^2 + (Z - Z^0)^2}}\,,
\end{equation}
where the equilibrium positions of the $X$ and $Y$ relative motions are zero by construction. Note that the harmonic terms from $V_\text{tr}$ simply decompose into oscillations of the center-of-mass and relative motion coordinates and do not interfere with the following harmonic expansion of $V_C$.

This expression can be expanded around equilibrium up to harmonic terms, where the zeroth order term is a constant irrelevant to the dynamics and the first order term is zero due to the equilibrium condition. The second order term is 
\begin{equation}
    \frac12\sum_{\alpha,\beta\in\{X,Y,Z\}}\frac{\partial^2 V_C}{\partial\alpha\partial\beta}\Big\rvert_{\substack{X=Y=0\\Z=Z^0}}~\alpha\beta\,.
\end{equation}
Evaluating this produces the expression
\begin{equation}
    \frac{e^2}{2(Z^0)^\frac52}\left(2Z^2-X^2-Y^2\right)=\frac{e^2}{2(z_1^0-z_2^0)^\frac52}\left(2(z_1-z_2)^2-(x_1-x_2)^2-(y_1-y_2)^2\right)\,.
\end{equation}
Thus the motion along the three axes is decoupled and the Hamiltonian for the motion along $x$ direction is 
\begin{equation} 
  H = \frac{1}{2m} \sum_i p_{i}^{2} +  \sum_{i} \frac{1}{2} m \omega_{x_i}^2 x_{i}^{2} + \frac{e^2}{2} \frac{\left(x_1-x_2\right)^{2}}{\left\vert z^{0}_{1} - z^{0}_{2}\right\vert^{3}}\,,
  \label{eqn:eqn2}
\end{equation}
which reproduces Eq.~(2) of the main text.

\begin{figure}[ht!]
\centering
 \includegraphics[height=5.25cm,width=13.5cm]{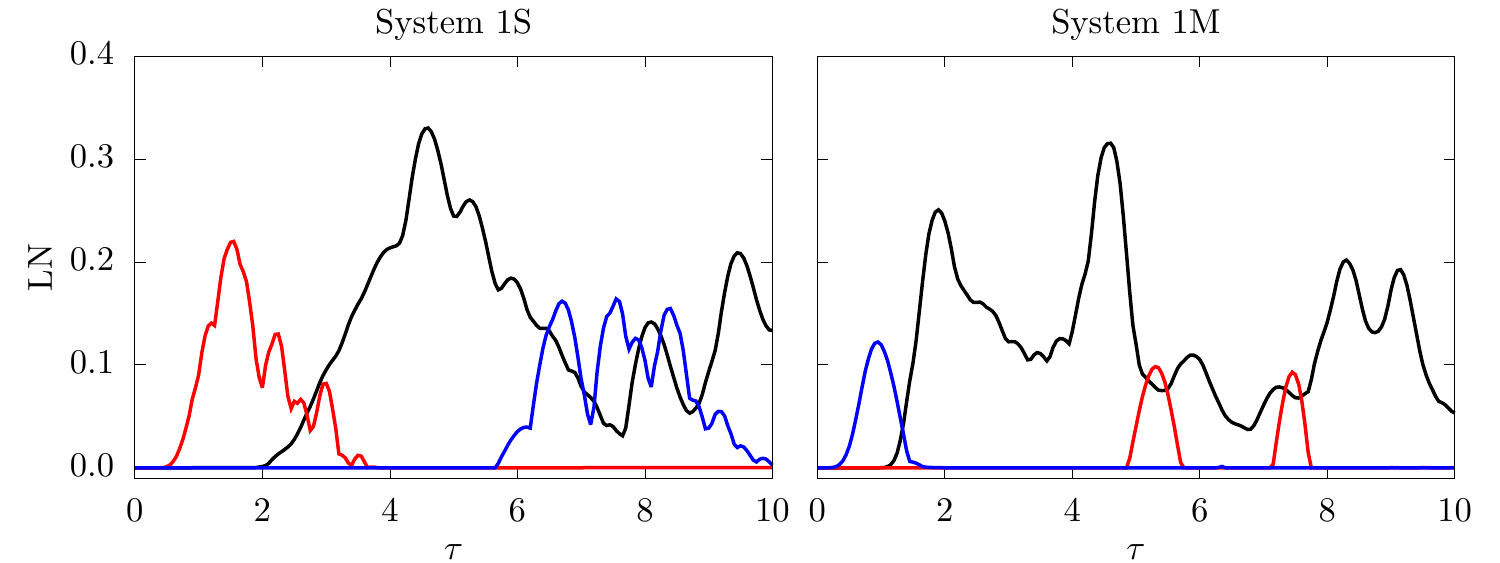}\vspace{-1ex}
\caption{The unitary time  evolution of LN between the oscillators (black), atom and oscillator 1 (red), and atom and oscillator 2 (blue) for both systems 1S (left) and 1M (right). The qubits in both the systems are initially kept in their respective ground states, while $\bar{n}_1 = 2,\, \bar{n}_2 = 0$.}
 \label{fig2}
\end{figure}

\section{Note on qubit-oscillator entanglement}
In the main text, for various model systems we discussed at length the behavior of oscillator-oscillator entanglement, characterised by logarithmic negativity (LN). In this section, we give details of our investigation on other bipartite entanglement between two subsystems that constitute each model. For instance, the qubit-oscillator entanglement for system $1S$ (Fig.~\ref{fig2}, left panel) roughly follows heuristic explanations of the entanglement generation. First, the oscillator with the noise becomes entangled with the qubit, then the nonclassical state of the oscillator induces entanglement between the oscillators through a linear interaction. Finally, the second oscillator becomes entangled with the atom which decreases the entanglement of the oscillators. There are residual correlations between the oscillators and the qubit, but they do not seem to be as significant as those created between the oscillators. System $1M$ is less easily explained in these terms (Fig.~\ref{fig2}, right panel). However, since the qubit and one of the oscillators begin in the ground state, the first dynamics must be the excitation of the qubit by the noisy oscillator. The qubit can then drive nonclassicality in the oscillator pair, with it first becoming visible in the logarithmic negativity between the qubit and the less noisy oscillator. Subsequently the two oscillators become entangled, mediated by the qubit. In the multiqubit case the qubit-qubit and qubit-oscillator entanglement is even less significant compared to that generated between the oscillators (Fig.~\ref{fig3}, left panel). 
  
\begin{figure}[ht!]
\centering
 \includegraphics[height=5.25cm,width=13.5cm]{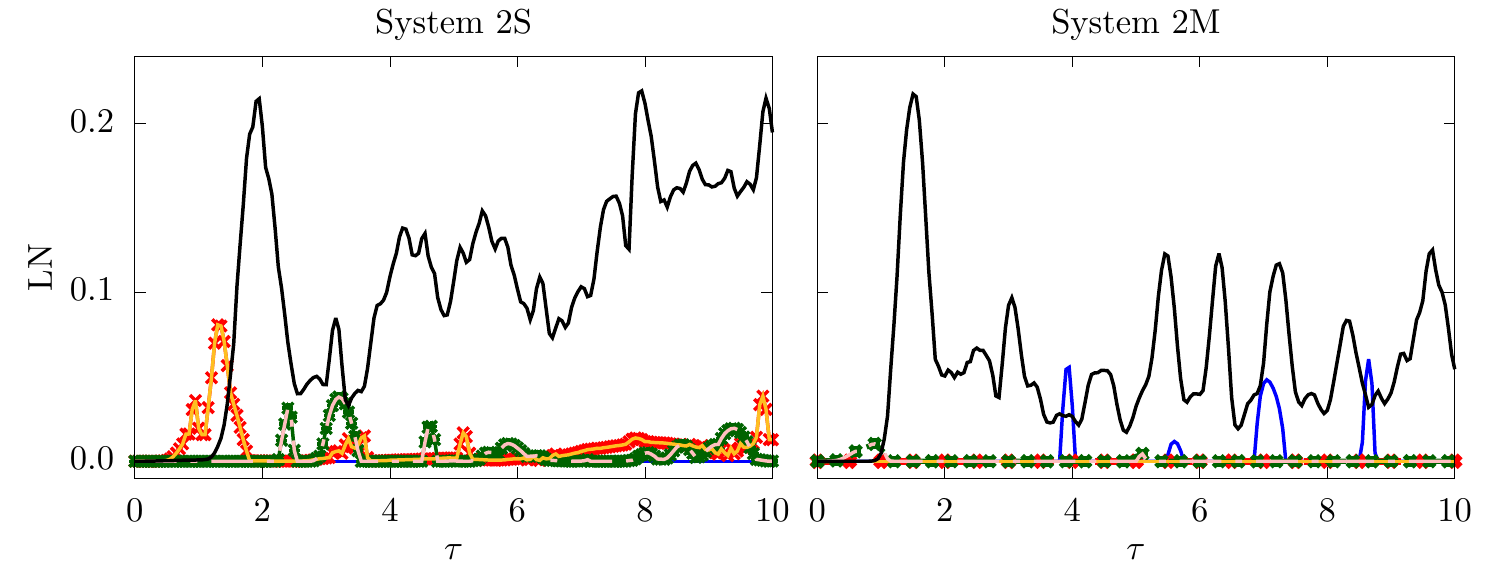}\vspace{-1ex}
\caption{The unitary time  evolution of LN between (a) oscillator-oscillator (black), atom-atom (blue),  atom 1-oscillator 1 (red), and atom 1-oscillator 2 (green), atom 2-oscillator 1 (yellow), and atom 2-oscillator 2 (pink) for both systems 2S (left) and 2M (right). The qubits in both the systems are initially kept the ground state, while $\bar{n}_1 = 2,\, \bar{n}_2 = 0$. Red and yellow produce the same dynamics. These can be directly traced back to the symmetry of the systems at the critical value. Similarly, green and pink. For system 1S atom-atom entanglement is extremely negligible, mostly zero.}
 \label{fig3}
\end{figure}

\section{Decoherence Effects}
In examining thermally induced entanglement among the vibrational modes of trapped ions we have assumed unitary dynamics. However small, there are still decoherence effects present in such systems. In this Section we give details of our numerical observations on the robustness of the entanglement preparation against such effects. In particular, we study the changes in the system dynamics due to the presence of qubit dephasing and other dissipative agents of open systems.
\begin{figure}[ht!]
\centering
\includegraphics[height=12.25cm,width=13.50cm]{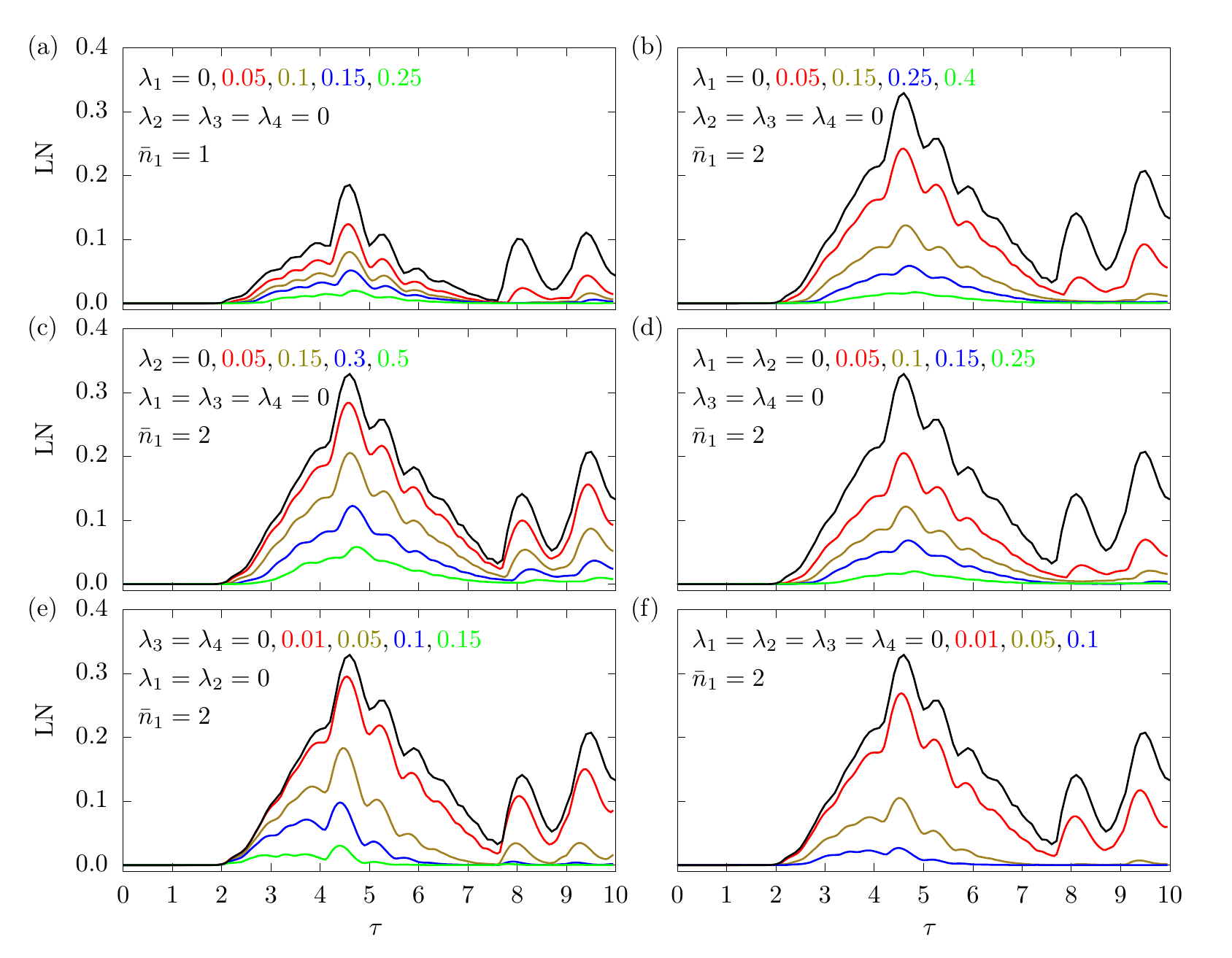} 
\vspace{-5.5ex}
\caption{Effect of decoherence and dissipation on the dynamics of LN for system $\mathrm{1S}$ with $p_{e}  = 0$, $\bar{n}_{2} = 0$ and $r_\mathrm{1S} = 0.18$.  A zero-temperature environment is assumed, i.e., $\bar{n}_{\text{th}} = 0$.  Various sources of dissipative effects are considered describing: (a)-(b) qubit dephasing for two different values of $\bar{n}_{1}$, (c) qubit dissipation, (d) dephasing and dissipation together, (e) mechanical decoherence of the vibrational modes, and  (f) combined effects of all the major dissipative effects.}
\label{fig:diss1}
\end{figure}
 Ignoring the contributions leading to a small renormalization of the system energy levels through the Lamb shift, the Lindblad master equation describing such environmental effects is  given by 
\begin{equation}
 \frac{d}{dt} \rho_{S}(t) = -i \left[H, \, \rho_{S}(t)\right] + \sum_{k} \left( L_{k}\, \rho_{S}(t)\, L_{k}^{\dagger} -  \tfrac{1}{2}\rho_{S}(t)\, L_{k}^{\dagger}\, L_{k} -  \tfrac{1}{2}L_{k}^{\dagger}\, L_{k}\, \rho_{S}(t)  \right).
\end{equation}
Here  $\rho_{S}(t)$ is the reduced density matrix of the system. The operators $L_{k} = \sqrt{\lambda}_{k} A_{k}$ are the Lindblad jump operators, and the environment couples to the system through the operators $A_{k}$ with coupling rates $\lambda_{k}$. We  focus solely on $H_{\mathrm{1S}} $ here as the conclusions drawn are generic to the remaining systems.

We first regard the environment as a zero-temperature bath. The various sources of decoherence and their associated Lindblad jump operators are qubit dephasing $A_1=\sigma_z$, qubit decoherence $A_2=\sigma_-$, and mechanical decoherence $A_3=a_1$ and $A_4=a_2$. Each have their associated decoherence rates, with $\lambda_3=\lambda_4$. The top panels of Fig.~\ref{fig:diss1} display the inhibition of entanglement generation due to qubit dephasing for two different values of $\bar{n}_{1}$ and for various dephasing rates $\gamma_1$. The effect of qubit relaxation alone on the dynamics is, however, less than that of dephasing  as can be seen by comparing Figs.~\ref{fig:diss1}(b) and (c). Fig.~\ref{fig:diss1}(d) shows the cumulative effect of qubit dephasing and relaxation. This suggests that the coherence of the qubit is an important factor in the entanglement dynamics. On the other hand Fig.~\ref{fig:diss1}(e) corresponds to the dissipative effect arising due to both the oscillators interacting with the environment. Finally, in Fig.~\ref{fig:diss1}(f) we see  the effect of all these environmental couplings. 
\begin{figure*}[ht!]
 \centering
 \includegraphics[height=8.25cm,width=13.5cm]{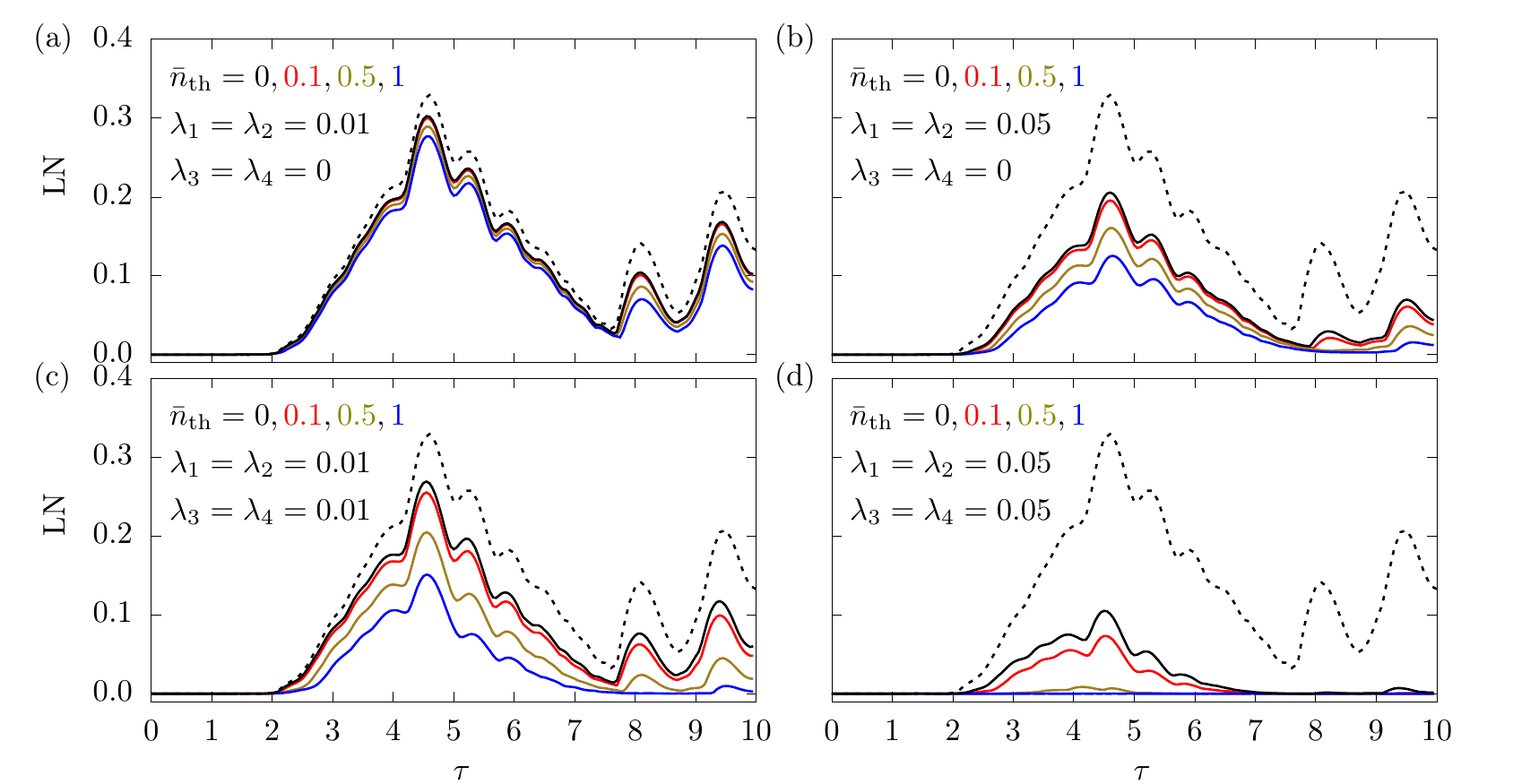} 
 \caption{Effect of thermal environment on the unitary dynamics of system I. The black dotted curves correspond to the isolated system dynamics. The top panel shows the effect of qubit dephasing and relaxation in a thermal environment whereas the bottom panel shows the effect of all the dissipation and decoherence effects on the dynamics.}
\label{fig:diss3}
\end{figure*}

Now we consider a non-zero temperature thermal environment. For this relatively simple analysis we set the average number of phonons, $\bar{n}_{\text{th}}$, to be the same for each dissipation channel. For relaxation and mechanical dissipation effects the decoherence rates are modified to the pairs 
$\sqrt{\lambda_{2}(1+\bar{n}_{\text{th}})}$ and $\sqrt{\lambda_{2}\, \bar{n}_{\text{th}}}$, and $\sqrt{\lambda_{i}(1+\bar{n}_{\text{th}})}$ and $\sqrt{\lambda_{i}\, \bar{n}_{\text{th}}}$ for $i=3,4$. The effect of increasing $\bar{n}_{\text{\text{th}}}$ on the dynamics is displayed in Fig.~\ref{fig:diss3} for various parameter regimes.

\section{COMPARISON: ENTANGLEMENT INDUCED BY COHERENT STATES}

An alternative to driving entanglement generation in passively coupled oscillators using classical correlations (thermal states) is to consider quantum coherence, i.e. quantum correlations internal to a vibrational mode. As mentioned, two of the resources for quantum technologies are entanglement and coherence, and hence it is relevant for autonomous generation of these resources to examine how they intersect. As an example, coherent states which are distributed differently in the Fock basis, show quantum coherence, and are pure states. Here we briefly consider the entanglement generation when the oscillator pair consists of a single coherent state and the ground state. 

\begin{figure*}[ht!]
 \centering
 \includegraphics[height=8.25cm,width=13.5cm]{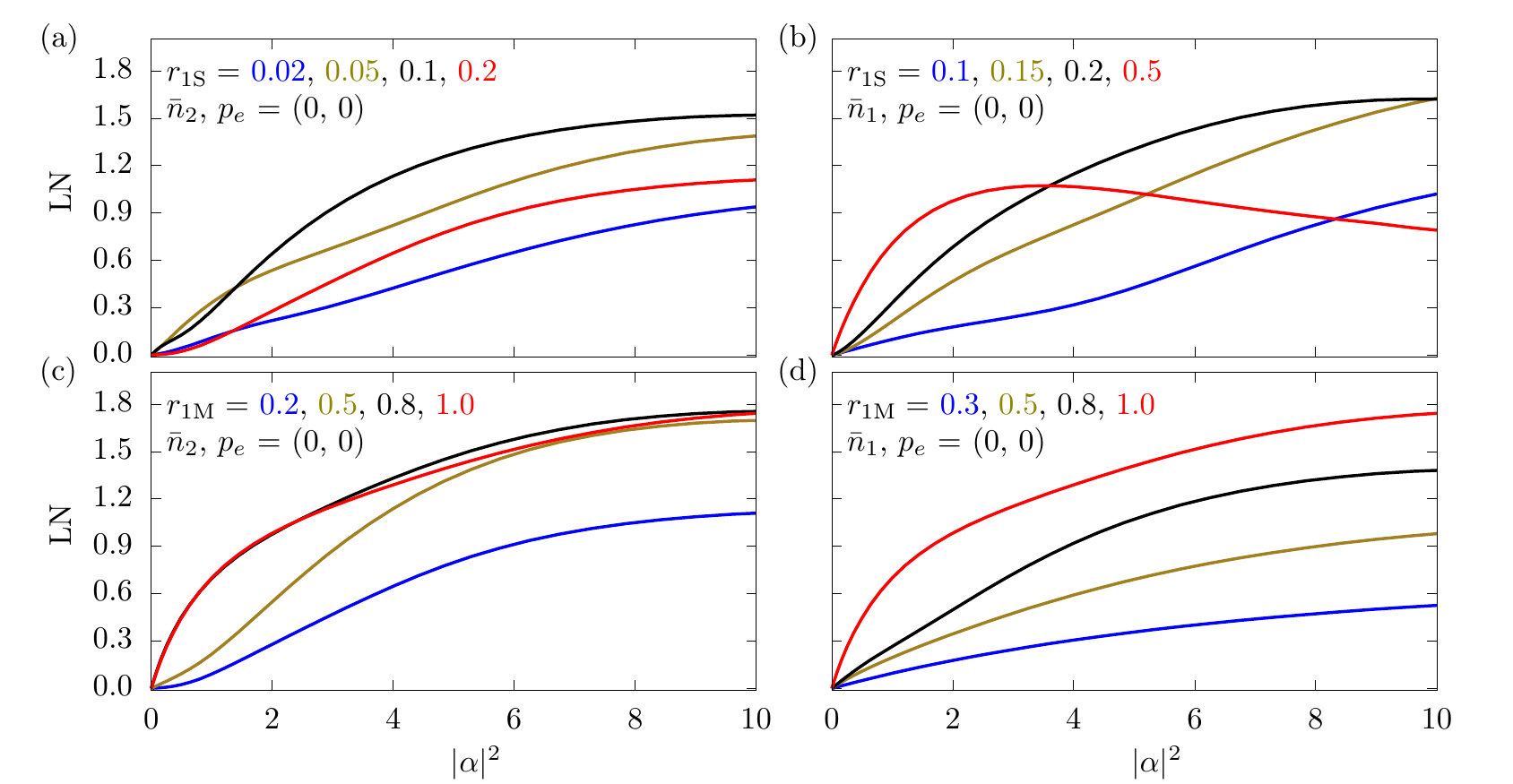} 
 \caption{Entanglement induced by a coherent state from the unitary dynamics governed by $H_{\mathrm{1S}} $ (top panel) and $H_{\mathrm{1M}} $ (bottom panel). The qubit is prepared in the ground state. The entanglement is significantly higher compared to the case of an initially thermal state of the oscillator. Left and right columns respectively correspond to oscillators 1 and 2 prepared as CS.}
\label{fig:cs}
\end{figure*}

The entanglement induced by a coherent state (CS) $\ket{\alpha}$ (expressed in the Fock basis as $\sum_{n} c_{n}\ket{n}$ where $c_{n} = e^{-\frac{1}{2}\vert\alpha\vert^{2}} \alpha^{n}/\sqrt{n!}$ and $\alpha \in \mathbb{C}$) of the oscillator is shown in Fig~\ref{fig:cs}. Again, the dynamics is unitary and the qubit is initially in the ground state. It is clear that the entanglement thus generated is significantly higher than that induced thermally for both $H_{\mathrm{1S}} $ and $H_{\mathrm{1M}} $. The critical value of the coupling $r_\mathrm{1S}$ is lower, although if $r_\mathrm{1S}$ is increased far away from the critical value the entanglement shows a decreasing trend with $\vert\alpha\vert^{2}$, after an initial increase. Once again, for  $H_{\mathrm{1M}} $ the symmetry in the system is borne out in the dynamics and maximum LN is achieved for $r_\mathrm{1M} = 1$. However, the entanglement generation is not monotonous with $|\alpha|^2$.

\begin{figure*}[ht!]
 \centering
 \includegraphics[height=8.25cm,width=13.5cm]{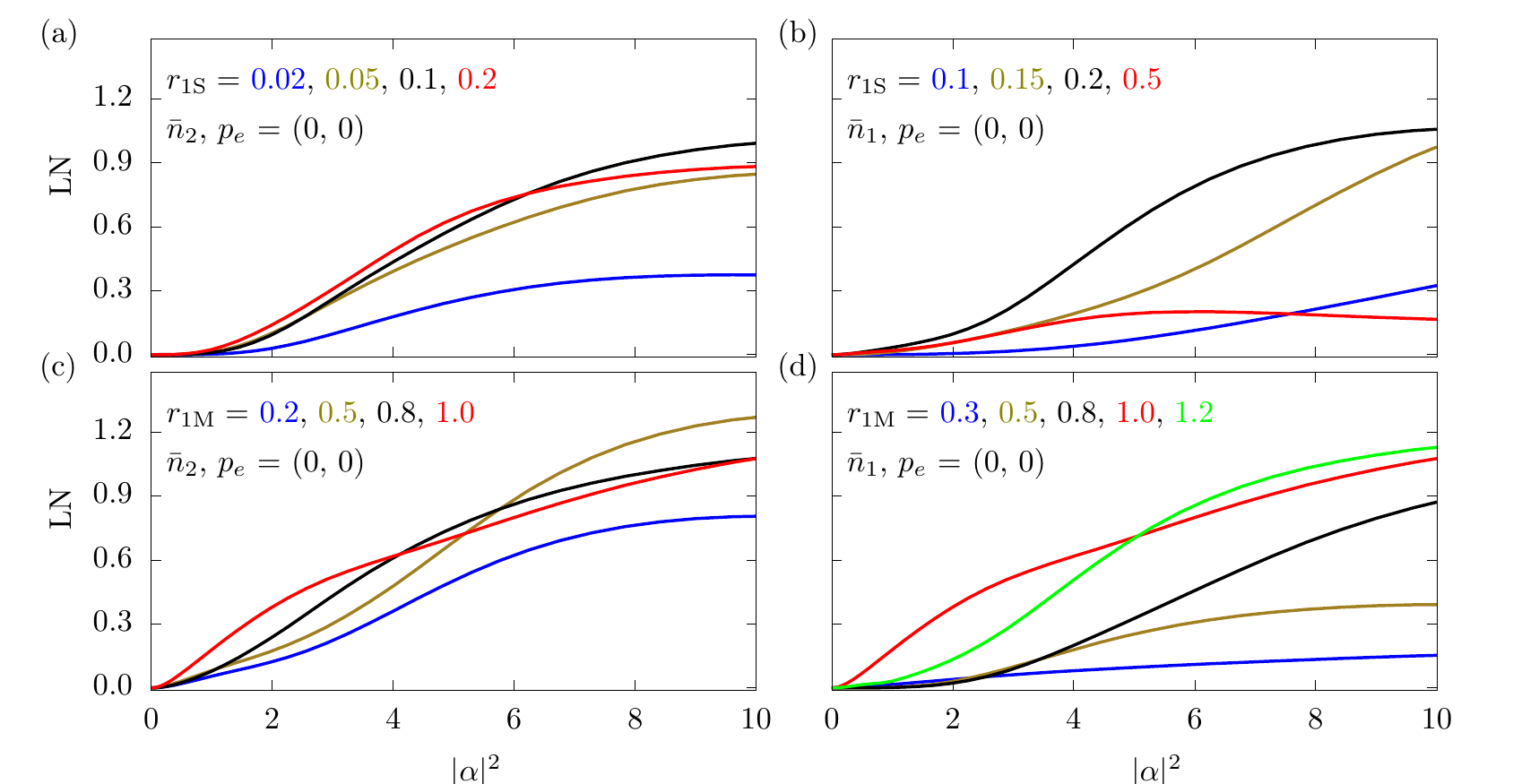} 
 \caption{Entanglement induced by a phase randomised coherent state from the unitary dynamics governed by $H_{\mathrm{1S}} $ (top panel) and $H_{\mathrm{1M}} $ (bottom panel). The qubit is prepared in the ground state. The entanglement is significantly higher compared to the case of an initially thermal state of the oscillator and less than that of standard coherent state. Left and right columns respectively correspond to oscillators 1 and 2 prepared as phase randomised coherent states.}
\label{fig:prcs}
\end{figure*}

\section{COMPARISON: ENTANGLEMENT INDUCED BY POISSONIAN NOISE}
To distinguish between the quantum coherence of the coherent states and merely the change in distribution from Boltzmann to Poisson we also examine the entanglement induced via phase randomised coherent states. When the phase of a coherent state is unknown it becomes a mixed state and the photon number follows a Poisson distribution with a mean of $\vert\alpha\vert^{2}$. Mathematically phase randomised coherent states are defined by
\begin{equation}
  \frac{1}{2\pi} \int_{0}^{2\pi} d\theta\, \ket{\alpha e^{i\theta}}\bra{\alpha e^{i\theta}} = \sum_{n=0}^{\infty} e^{-\vert\alpha\vert^{2}}\frac{\vert\alpha\vert^{2n}}{n!}\ket{n}\bra{n}.
\end{equation}
Therefore the number states are incoherent with each other.

The entanglement induced by such a state of the oscillator is shown in Fig~\ref{fig:prcs}. Again, the dynamics is unitary and the qubit is initially in the ground state. It is clear that the entanglement generated in this case is significantly higher than that induced thermally for both $H_{\mathrm{1S}} $ and $H_{\mathrm{1M}} $ but less than that of the coherent states. Again, the entanglement generation is not monotonous with $|\alpha|^2$.

\section{Comparison: Three Qubit Case}
In the main text we have considered a protocol wherein the passively coupled oscillators are prepared in thermal equilibrium, and this constitutes the resource (alongside coherent coupling with a two-level system) for inducing entanglement. Decreasing the purity of the qubit from the ground state negatively affects entanglement generation. However, alternative strategies can be devised. We commented on them briefly in the discussion, and here we give a fuller treatment. 
\begin{figure}[ht!]
\centering
\includegraphics[height=5.25cm,width=13.5cm]{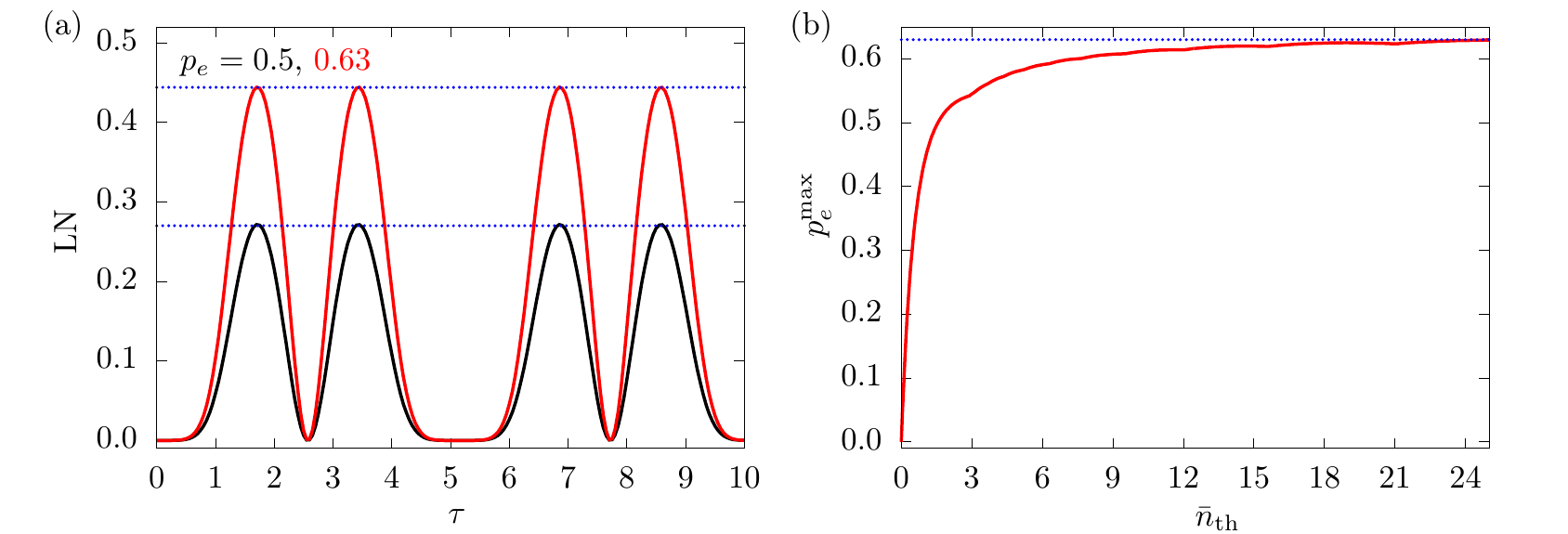} 
\caption{The alternative initial state preparation, wherein the qubit is in the thermal state and both the oscillators are in their respective ground states, leads to an effective $3-$qubit Hamiltonian given by Eq.~\ref{eq:3qubit}, for both Hamiltonians $H_{\mathrm{1S}}$ and $H_{\mathrm{1M}}$. Here we excite qubit 1 to a thermal state and evaluate the LN between qubits 2 and 3. We consider two different choices of the initial qubit state: (i) a maximally thermal qubit with $p_{e}=0.5$ (black curve in (a)) and (ii) $p_{e}=0.63$ (red curve in (a)). The latter is obtained by coupling a ground state qubit to a thermal oscillator through a coherent JC interaction. The unitary dynamics produces population inversion and in such a setting $p_{e}$ saturates to the value 0.63 (red curve in (b)). The blue dotted lines in (a) correspond to $\text{LN} = 0.27$ and $0.44$ respectively, whereas the blue dotted line in (b) corresponds to $p_{e}=0.63$.}
\label{fig:3qubit}
\end{figure}
\begin{figure}[ht!]
\centering
\includegraphics[height=8.25cm,width=13.5cm]{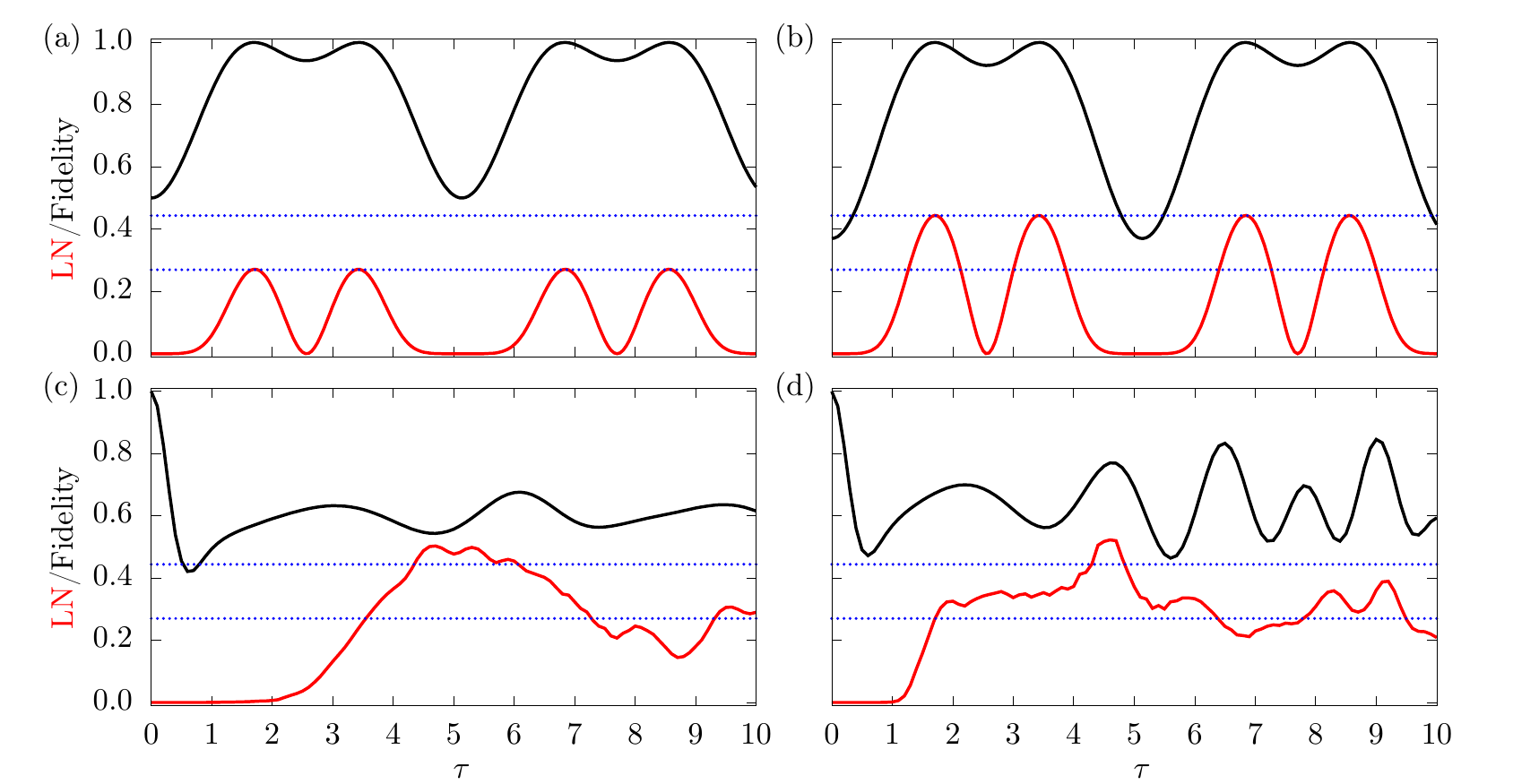} 
\caption{Qubit fidelity analysis for an effective $3-$qubit Hamiltonian given by Eq.\eqref{eq:3qubit} (top panel), and for Hamiltonians $H_{\mathrm{1S}}$ and $H_{\mathrm{1M}}$ ((c) and (d) respectively) corresponding to initial thermal oscillators and ground state qubit (similar to Figs.
1(a-b) with $\bar{n}_1 = 5$ and $\bar{n}_{2}=0$). In (a) we consider a maximally thermal qubit with $p_{e}=0.5$ and in (b) we choose $p_{e}=0.63$. In all these cases the black curves correspond to the qubit fidelity with respect to the ground state and the red curves to LN of the joint oscillators state.  The blue dotted lines correspond to $\text{LN} = 0.27$ and $0.44$ respectively.}
\label{fig:fidelity}
\end{figure}

One such alternative strategy is to consider a thermal equilibrium qubit and ground state oscillators. In this scenario there is a maximum of one excitation due to the nonzero probability of finding the qubit in the excited state. Thus, the full system may be well-approximated by a three-qubit system, where two of the qubits are prepared in their respective ground states and one in a thermal state. The models $\mathrm{1S}$ and $\mathrm{1M}$ in this approximation are equivalent and are described by the following Hamiltonian
\begin{equation}
    H_{3Q}=\kappa_1(\sigma_+^1\sigma_-^2+\text{h.c.})+\kappa_2(\sigma_+^2\sigma_-^3+\text{h.c.})\,,
    \label{eq:3qubit}
\end{equation}
where numerical superscripts label the qubits. The label 1 always refers to the thermal qubit while 2 and 3 denote the proxies for the oscillator pair. For thermal qubits, $\text{max}(p_e)=\frac12$ and the maximum available entanglement generated among the remaining qubits is $\text{LN}\approx0.27$ (black curve in Fig.~\ref{fig:3qubit}(a)). This contrasts with the nonequilibrium state $p_e=1$ which prepares maximally entangled pairs of qubits ($\text{LN} = 1$). Both system $\mathrm{1S}$ and system $\mathrm{1M}$ are sufficient to surpass the entanglement induced by a thermal qubit. 

The possibility of approaching a maximally entangled state via $p_e > \frac12$, i.e. population inversion, motivates us to attempt to use thermal equilibrium states to generate such qubit states. By coupling a ground state qubit to a thermal oscillator, one may prepare a nonequilibrium qubit with $p_e\approx0.63$, using only equilibrium resources (red curve in Fig.~\ref{fig:3qubit}(b)). If this state is selected as the initial state in the three-qubit approximation, the maximum achievable entanglement is enhanced to $\text{LN} \approx 0.44$. This is below the level of entanglement achievable with $H_{\mathrm{1S}}$. With $H_{\mathrm{1M}}$ there is also the capacity to cross this threshold, but it occurs for the second entanglement peak, which we do not analyse. 
\section{Note on Numerical Methods}

Our simulations take place in a truncated Hilbert space characterised by a Fock dimension $n$, which characterises the size of each subspace associated with a vibrational mode. To ensure stability and accuracy we adhere to an error threshold of $10^{-6}$, much higher than the precision typically available in experiment.

\begin{figure}[ht!]
\centering
\includegraphics[height=5.5cm,width=8.5cm]{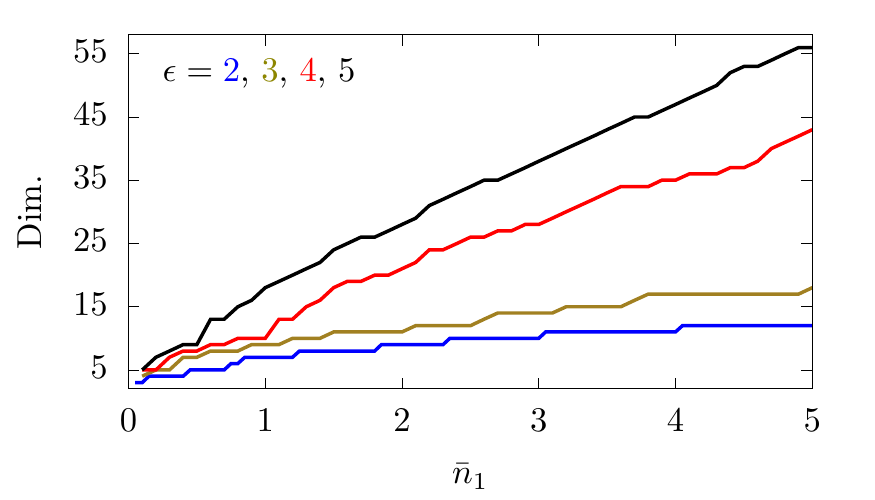} 
\caption{The truncation dimension of the oscillator Hilbert spaces increases with rising $\bar{n}_1$, if the error threshold is held constant.}
\label{fig:dim_n1_h1}
\end{figure}

Fig.~\ref{fig:dim_n1_h1} shows how the truncation dimension $n$ scales with the error threshold. The error threshold parameter is referred to as the power scaling $10^{-\epsilon}$. As expected higher accuracy requires higher dimensions. In order to maintain this accuracy with increasing $\bar{n}_{1}$ the dimension must rise quickly. The figure presented is generated for system I (with $p_{e} = \bar{n}_{2} = 0$ and $r_\mathrm{1S} = 1$), but the result is generic and holds qualitatively for all other systems.

\section{Importance of the Ground State}

In the main text we emphasised that the initial state of the qubit being the ground state is important for thermally induced entanglement. Here we expand on this by examining the state of the qubit throughout the entanglement generation process. The most clear-cut example occurs for the approximation in Section~5.
The fidelity of the qubit with the ground state, alongside the evolution of the entanglement of the remaining pair of qubits is shown in Fig~\ref{fig:fidelity}. For this unitary system, the behaviour is simple and oscillatory, showing collapses and revivals of the entanglement and of the excitation of the thermal qubit. Despite this simple behaviour, a full collapse and revival of the entanglement occurs while the thermal qubit stays close to the ground state, eluding simple explanations based on the transfer of a single excitation. 

More complicated behaviour occurs for the main protocol of the article and is shown in Fig.~\ref{fig:fidelity}. Here the correlation of the ground state of the qubit with the entangled states is much weaker. However, it is clear that the qubit undergoes some evolution before the entanglement emerges, reflecting the nontrivial dynamics generating the entanglement.

\section{Early Entanglement Dynamics}
\begin{figure}[ht!]
\centering
\includegraphics[height=5.0cm,width=13.5cm]{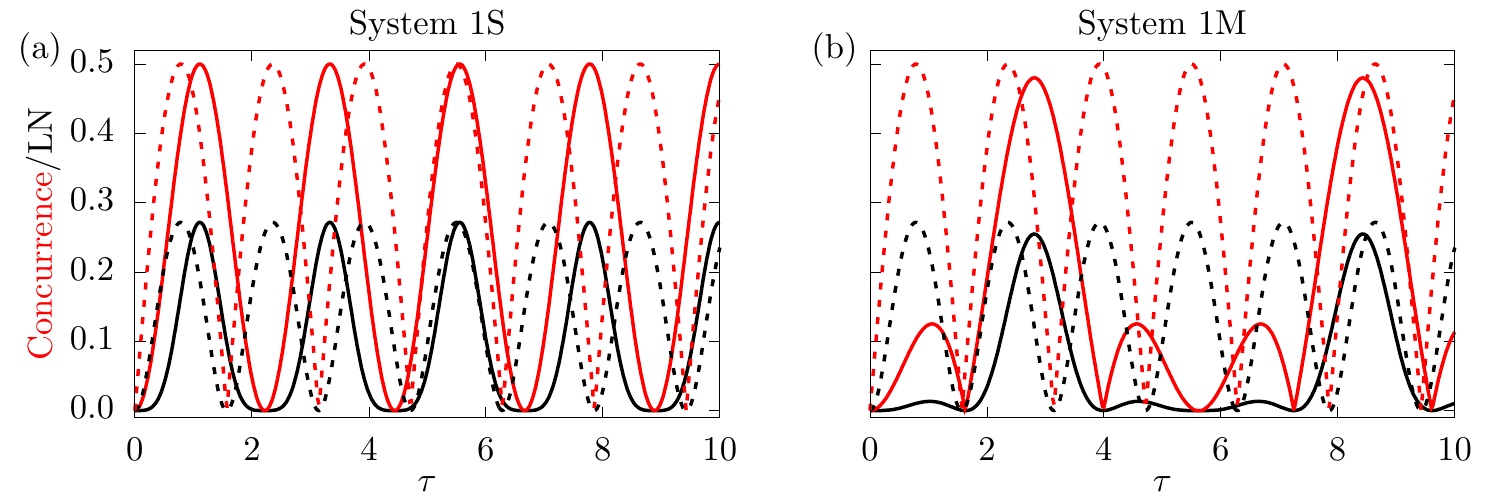} 
\caption{In the three qubit analogy of system 1S (left) and 1M (right) the dynamics of concurrence (red solid curves) and LN (black solid curves) are shown as a function of time.  $p_e = 0.5$ in both cases. $r_{\mathrm{1S}} = 1$ and  $r_{\mathrm{1M}} = 0.5$. Left: Qubit 2 is thermally excited and the correlations are evaluated between qubits 2 and 3. Right: Qubit 1 is excited and the correlations are evaluated between qubits 1 and 3. Both concurrence and LN are compared (dashed curves) with the  corresponding two qubit system with $H=\kappa(\sigma_{1+}\sigma_{2-} + \sigma_{1-}\sigma_{2+})$ with qubit 1 initially in the maximally thermal state and qubit 2 in the ground state.}
\label{fig:conc}
\end{figure}

In the main text we observe that the logarithmic negativity is zero for the early parts of the dynamics. LN is a sufficient condition of entanglement and therefore can be zero even when the state is entangled.  For such large-dimensional systems it is difficult to evaluate this region more clearly. However we can compare the entanglement dynamics of the LN with those of the concurrence (in the three qubits analogy, Eq.~(\ref{eq:3qubit})). The concurrence is shown in Fig.~\ref{fig:conc} against the LN. The concurrence has its major peaks synchronised with the LN, however for system 1M, if we were to follow our principle of taking the first peak of the entanglement dynamics for analysis of thermally induced effects we would miss the possibility of large entanglement that the LN captures.

\begin{figure}[ht!]
\centering
\includegraphics[height=5.0cm,width=7.0cm]{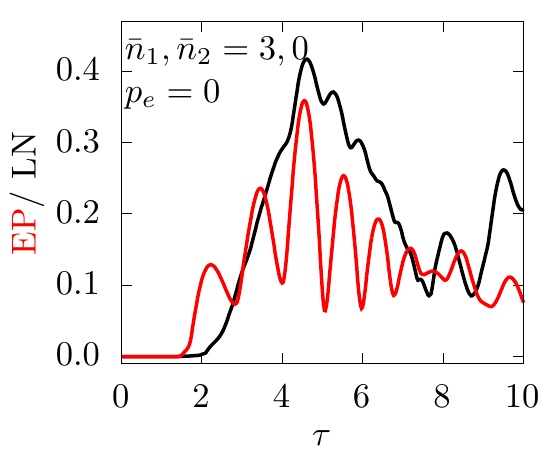} 
\caption{The entanglement potential (EP) of the thermal oscillator in the absence of the linear BS coupling is contrasted with the LN between the oscillators with the BS coupling on.  The qubit is in the ground state, while $\bar{n}_1 = 3,\, \bar{n}_2 = 0$.}
\label{fig:ep}
\end{figure}

\section{Entanglement potential and LN}
The nonclassicality of a state $\rho$ capable of being converted to entanglement can be characterised through its entanglement potential (EP), which characterises the state's capacity to become entangled via passive interactions with the ground state of an oscillator. This is quantified by the logarithmic negativity (LN), of the following state:
\begin{equation}
 \displaystyle \rho_{\mathcal{E}}=U_\text{BS}^\dagger\,\left(\rho\otimes\ket{0}\bra{0}\right)U_ \text{BS}, 
 \label{eqn:bs}
\end{equation}
where $U_{\text{BS}} = e^{\frac{\pi}{4}(c^{\dagger} a - c a^{\dagger})}$ and $c$ is an auxiliary mode prepared in the ground state.

 The EP for a qubit coupled to a thermal oscillator is shown in red in Fig.~\ref{fig:ep}. In contrast, the black curve shows the corresponding LN (see also Fig.~2 of the main text) when the thermal oscillator is simultaneously coupled to a ground state oscillator. The EP of the thermal oscillator increases before the LN of the full system, supporting our interpretation that first the qubit must drive the thermal oscillator to a nonclassical state before it can entangle with the second oscillator. However the LN eventually surpasses the EP. This suggests that the back action from the qubit continues to drive the nonclassicality of the thermal oscillator while it generates entanglement with the second oscillator.

\end{document}